\documentclass[manuscript,screen]{acmart}
\AtBeginDocument{%
  }

\usepackage[utf8]{inputenc}
\usepackage{algorithm}
\usepackage[noend]{algpseudocode}
\usepackage[normalem]{ulem}
\usepackage{amsmath}
\usepackage{amsthm}
\newtheorem{remark}{Remark}
\theoremstyle{definition}
\newtheorem{definition}{Definition}
\newtheorem{example}{Example}

\usepackage{natbib}
\usepackage{tikz}
\usepackage{tikzpeople}
\usetikzlibrary{calc}
\usetikzlibrary{decorations.pathreplacing}
\usetikzlibrary{decorations.pathmorphing}
\usetikzlibrary{shapes.geometric}
\usepackage{xcolor}
\usepackage{listings}
\usepackage{pgfplots}
\pgfplotsset{compat=1.18}

\algrenewcommand\alglinenumber[1]{\scriptsize #1}

\errorcontextlines\maxdimen

\RequirePackage{tikz}
\RequirePackage{zref-abspage}
\RequirePackage{zref-user}
\RequirePackage{tikz}
\RequirePackage{atbegshi}
\usetikzlibrary{calc}
\RequirePackage{tikzpagenodes}
\RequirePackage{etoolbox}

\makeatletter
\newcommand*{\algrule}[1][\algorithmicindent]{%
  \hspace*{.2em}%
  \vrule %
  \hspace*{\dimexpr#1-.2em-.4pt}%
}

\newcommand{\StatePar}[1]{%
  \State\parbox[t]{\dimexpr\linewidth-\ALG@thistlm}{\strut #1\strut}%
}
\renewcommand{\ALG@beginalgorithmic}{\offinterlineskip}%

\newcount\ALG@printindent@tempcnta
\def\ALG@printindent{%
  \ifnum \theALG@nested > 0%
    \ifx\ALG@text\ALG@x@notext%
    \else
      \unskip
      \ALG@printindent@tempcnta=1
      \loop
        \algrule[\csname ALG@ind@\the\ALG@printindent@tempcnta\endcsname]%
        \advance \ALG@printindent@tempcnta 1
        \ifnum \ALG@printindent@tempcnta<\numexpr\theALG@nested+1\relax
      \repeat
        \fi
    \fi
}
\patchcmd{\ALG@doentity}{\noindent\hskip\ALG@tlm}{\ALG@printindent}{}{\errmessage{failed to patch}}
\makeatother

\algrenewcommand\algorithmicend{\strut\textbf{end}}
\algrenewcommand\algorithmicdo{\strut\textbf{do}}
\algrenewcommand\algorithmicwhile{\strut\textbf{while}}
\algrenewcommand\algorithmicfor{\strut\textbf{for}}
\algrenewcommand\algorithmicforall{\strut\textbf{for all}}
\algrenewcommand\algorithmicloop{\strut\textbf{loop}}
\algrenewcommand\algorithmicrepeat{\strut\textbf{repeat}}
\algrenewcommand\algorithmicuntil{\strut\textbf{until}}
\algrenewcommand\algorithmicprocedure{\strut\textbf{procedure}}
\algrenewcommand\algorithmicfunction{\strut\textbf{function}}
\algrenewcommand\algorithmicif{\strut\textbf{if}}
\algrenewcommand\algorithmicthen{\strut\textbf{then}}
\algrenewcommand\algorithmicelse{\strut\textbf{else}}

\let\oldState\State
\renewcommand{\State}{\oldState\strut}

\definecolor{codegreen}{rgb}{0,0.6,0}
\definecolor{codegray}{rgb}{0.5,0.5,0.5}
\definecolor{codeblue}{rgb}{0,0,0.5}
\definecolor{codepurple}{rgb}{0.58,0,0.82}
\definecolor{darkgreen}{rgb}{0.0, 0.2, 0.13}

\DeclareFixedFont{\ttb}{T1}{txtt}{bx}{n}{8pt} %
\DeclareFixedFont{\ttm}{T1}{txtt}{m}{n}{8pt}  %
\lstdefinestyle{pythonstyle}{
  language=python,
  deletekeywords={int},
  commentstyle=\color{codegreen},
  keywordstyle=\ttb\color{codeblue},
  numberstyle=\tiny\color{codegray},
  stringstyle=\ttm\color{codepurple},
  basicstyle=\ttm,
  breakatwhitespace=false,
  breaklines=true,
  captionpos=b,
  keepspaces=true,
  showspaces=false,
  showstringspaces=false,
  showtabs=false,
  tabsize=2,
  escapeinside={(*@}{@*)}
}

\lstdefinestyle{bnfstyle}{
  language=Python,
  deletekeywords={list},
  basicstyle=\ttm,
  keywordstyle=\ttm,
  breakatwhitespace=false,
  breaklines=true,
  captionpos=b,
  keepspaces=true,
  showspaces=false,
  showstringspaces=false,
  showtabs=false,
  tabsize=2,
  escapeinside={(*@}{@*)}
}

\lstdefinestyle{cstyle}{
  language=c,
  deletekeywords={int},
  commentstyle=\color{codegreen},
  keywordstyle=\ttb\color{codeblue},
  numberstyle=\tiny\color{codegray},
  stringstyle=\color{codepurple},
  basicstyle=\ttm,
  breakatwhitespace=false,
  breaklines=true,
  captionpos=b,
  keepspaces=true,
  showspaces=false,
  showstringspaces=false,
  showtabs=false,
  tabsize=2,
  escapeinside={(*@}{@*)}
}

\usepackage{graphicx}
\usepackage{caption}
\usepackage{enumitem}
\usepackage{subcaption}

\begin{document}

\title{Canonicalization of Batched Einstein Summations for Tuning Retrieval}
\author{Kaushik Kulkarni}
\email{kgk2@illinois.edu}
\orcid{0009-0001-0645-2169}
\affiliation{%
  \institution{Siebel School of Computing and Data Science, University of Illinois at Urbana-Champaign}
  \city{Urbana}
  \state{Illinois}
  \country{USA}
}
\author{Andreas Kl\"ockner}
\email{andreask@illinois.edu}
\orcid{0000-0003-1228-519X}
\affiliation{%
  \institution{Siebel School of Computing and Data Science, University of Illinois at Urbana-Champaign}
  \city{Urbana}
  \state{Illinois}
  \country{USA}
}

\begin{abstract}
  We present an algorithm for normalizing \emph{Batched Einstein Summation}
  expressions by mapping mathematically equivalent formulations to a unique
  normal form.  Batches of einsums with the same Einstein notation that exhibit
  substantial data reuse appear frequently in finite element methods (FEM),
  numerical linear algebra, and computational chemistry. To effectively exploit
  this temporal locality for high performance, we consider groups of einsums in
  batched form.

  Representations of equivalent batched einsums may differ due to index
  renaming, permutations within the batch, and, due to the commutativity and
  associativity of multiplication operation. The lack of a canonical
  representation hinders the reuse of optimization and tuning knowledge in
  software systems. To this end, we develop a novel encoding of batched einsums
  as colored graphs and apply graph canonicalization to derive a normal form.

  In addition to the canonicalization algorithm, we propose a representation of
  einsums using functional array operands and provide a strategy to transfer
  transformations operating on the normal form to \emph{functional batched
  einsums} that exhibit the same normal form; crucial for fusing surrounding
  computations for memory bound einsums.  We evaluate our approach against JAX,
  and observe a geomean speedup of $4.7\times$ for einsums from the TCCG
  benchmark suite and an FEM solver.
\end{abstract}

    


\maketitle

\section{Introduction}\label{sec:intro_mat_avoid_abstraction}
Linear Algebra subprograms serve as an abstraction layer between performance
engineers and computational scientists. The role of a performance engineer is
to provide optimized kernels tailored to specific hardware architectures, while
a computational scientist implements their application by composing these
kernels. Well-known examples for such abstractions include
\textsc{BLAS}~\cite{lawson1979basic} for primitives on dense linear operator
applications, \textsc{LAPACK}~\cite{anderson1999lapack} for solving linear
systems, \textsc{PETSc}~\cite{balay2001petsc} for iterative linear / non-linear
solvers, among others. By relying on such abstractions, computational
scientists can focus on formulating their applications while leveraging the
optimized routines provided by the implementation of such systems to achieve
performant execution on the underlying hardware systems.

A key operation observed across many scientific domains is the multilinear
algebra operator application. Such operators generalize inner and outer products
to multidimensional arrays and appear in a wide range of contexts. For example,
expressions of the form $\sum_{x j} J[x,r,e] D[x,i,j]u[e,j]$ appear during matrix-free
assembly in Finite Element Methods (FEM)~\cite{kirby2006optimizing}, expressions
of the form $\sum_{c d} g[i,j,c,d]t[c,d,a,b]$ appear in cluster operators in
Computational Chemistry~\cite{solomonik2014massively}, and expressions of the
form $\sum_{j_1 j_2} G_1[i_1,j_1] G_2[i_2,j_2]X[j_1,j_2]$ appear in Tensor Train
Layers in Neural Networks~\cite{alexander_2015_tensortrain}, etc.  {\AA}hlander
et al.~\cite{aahlander2002einstein} proposed a Domain-Specific Language (DSL)
that utilizes the Einstein Summation (``einsum'') notation as an abstraction to
offer a terse syntax for expressing such multidimensional contractions.  This
notation has since been adopted by a number of systems that support operations
on arrays, including \textsc{NumPy}~\cite{harris2020array},
\textsc{JAX}~\cite{frostig2018compiling},
\textsc{Cyclops}~\cite{solomonik2013cyclops}, and
\textsc{Deinsum}~\cite{ziogas2022deinsum}.

In large-scale applications, einsums rarely occur in isolation; instead, they
are often evaluated in groups with repeated access to the same data. For
example, in FEM applications, each derivative computation corresponds to the
same einsum expression applied to different field vectors, while certain
operands, such as geometric factors, remain unchanged. To represent this data
reuse, we find it critical to introduce a computational primitive corresponding
to a collection (unordered) of einsums sharing the same Einstein notation but
applied to different array operands. We use the term ``batched einsum'' for this
concept.

The ubiquity of batched einsums in scientific workloads makes it essential to
optimize their execution speed for available target hardware. Utilizing
state-of-the-art compilation techniques for einsums to lower batched einsums
often face two key drawbacks.  Firstly, these approaches tend to lower einsums
into a program by obtaining an optimal contraction path and subsequently relying
on a tensor contraction
code-generator~\cite{hirata2003tensor,bientinesi2018ttgt,kim2019code,matthews2018high}.
This introduces overheads from the allocation and initialization of intermediate
arrays, referred to as materialization overheads. Secondly, they do not fuse the
data accesses across einsum invocations leading to further overheads from
DRAM-to-register reads.  For these reasons, we take a contrasting approach and
\uline{design a software system that maintains a tabulation of high-performant
batched einsums to be reused in practical programs}.

We refer to the software systems that provide a tailored implementation for an
instance of a computational primitive as adopting a \emph{tabulation-based}
approach. Encoding batched einsums within such a table for retrieval poses a key
challenge due to the non-canonical nature of their representation. This can lead
to inefficiencies, with multiple table entries corresponding to mathematically
equivalent computations. We illustrate this in
Figure~\ref{fig:demo_sameish_einsums}, where two syntactically different batched
einsums can share the same computational implementation after appropriate
renaming. In Section~\ref{sec:isomorphic_batched_einsum}, we formalize these
equivalent classes of batched einsums by considering duplicates that may arise
from index renaming, permutations within a batch, and permutations of operands
within an einsum. Additionally, we have developed a \textit{canonicalization
algorithm} that maps a batched einsum expression to a canonical form.  This
algorithm ensures that equivalent batched einsum expressions are mapped to a
unique representation.

\begin{figure}[t]
  \begin{minipage}{0.4\linewidth}
    \fbox{
      \begin{minipage}{\linewidth}
        \begin{align*}
          R_1[i, j, k, l]  \gets  \sum_{m, n} A[m, n, j] B[m, n, i] C[k, l, i] \\
          R_2[i, j, k, l]  \gets  \sum_{m, n} D[m, n, j] E[m, n, i] A[k, l, i] \\
          R_3[i, j, k, l]  \gets  \sum_{m, n} B[m, n, j] D[m, n, i] F[k, l, i]
        \end{align*}
    \end{minipage}}
    \subcaption{Batched einsum $e_1$.}
  \end{minipage}\hspace{0.03\linewidth}
  \begin{minipage}{0.55\linewidth}
    \fbox{
      \begin{minipage}{\linewidth}
        \begin{align*}
          R_1'[n', m', k', j']  \gets  \sum_{i', l'} A'[l', i', n']D'[k',
          j', n']E'[l', i', m']\\
          R_2'[n', m', k', j']  \gets  \sum_{i', l'} E'[l', i', n']B'[k',
          j', n']C'[l', i', m']\\
          R_3'[n', m', k', j']  \gets  \sum_{i', l'} C'[l', i', n']F'[k',
          j', n']D'[l', i', m']
        \end{align*}
    \end{minipage}}
    \subcaption{Batched einsum $e_2$.}
  \end{minipage}
  \caption{Although $e_1$ and $e_2$ appear distinct set of expressions, the
    implementation of $e_1$ can be utilized for computing $e_2$ provided the
    following substitutions are applied to it: $R_2 \to R_1'$, $R_3 \to R_2'$,
    $R_1 \to R_3'$, $i \to n'$, $j \to m'$, $k \to k'$, $l \to j'$, $m \to l'$, $n
    \to i'$, $A \to D'$, $B \to C'$, $C \to F'$, $D \to E'$, $E \to A'$, and $F
  \to B'$.}\label{fig:demo_sameish_einsums}
\end{figure}

A key drawback of systems that adopt a tabulation-based approach for optimized
subprograms, like BLAS, is the requirement of the operands of the computational
primitive to be materialized. These materialization overheads limit the
application's computational throughput, particularly when the primitive itself
is memory bound. We illustrate this in Algorithm~\ref{alg:intro_blas_overheads},
where the low arithmetic intensity of matrix-vector product makes it is
generally profitable to fuse the computation of $q_1$ with the \textsc{BLAS}'
\textsc{GEMV}\footnote{\textsc{GEMV} corresponds to a dense matrix-vector
product operation} call. Such unmaterialized array expressions appear as
operands to an einsum in many scientific applications. For example, in the
computation of the 1-D Coulomb interaction matrix which takes in an elementwise
reciprocal of the distance matrix as an operand.  Such materialization overheads
also appear in high-level compilers that rely on libraries of optimized
subprograms, as they are often required to aggressively materialize intermediate
values in order to invoke these subprograms. This can be observed in lazy array
frameworks such as JAX and \textsc{PyTorch}~\cite{paszke2019pytorch}, which rely
on dense linear algebra kernels provided by \textsc{cuBLAS}. In certain cases,
this leads to inefficient generated binaries due to kernel boundaries that
inhibit fusion.

We address these materialization-induced penalties through two key ideas. First,
the computational scientist interacts with a functional form of the batched
einsum expression, in which the operands of the einsum are specified as
$\lambda$-expressions. Algorithm ~\ref{alg:functional_intro_blas_overheads}
illustrates how a computational scientist can express a computation as a
\emph{functional batched einsum}.  Second, we maintain a database that records
the code transformations required to optimize a \emph{functional batched einsum}
when each $\lambda$ operand corresponds to a C-style multidimensional array.
This design enables the computational scientist to reuse optimizations developed
for idealized batched einsums to transform programs that structurally resemble a
batched einsum.  Such transfer of code optimizations is becoming increasingly
important as modern accelerators, such as the NVIDIA H100, NVIDIA V100, and AMD
MI250X, exhibit very high saturation arithmetic intensity. Consequently, for
workloads that are memory-bound, performing additional floating-point operations
on data already resident on the processing units would have virtually no impact
on the execution time.

\begin{figure*}[ht]
  \centering

  \begin{minipage}[t]{0.47\textwidth}
    \algrenewcommand{\alglinenumber}[1]{\color{magenta}\footnotesize#1}
    \begin{algorithm}[H]
      \caption{Pseudocode for computing $\nabla\cdot(\sin(q))$ using FEM.}
      \label{alg:intro_blas_overheads}
      \begin{algorithmic}[1]
        \Procedure{FEMOperator}{$D$, $q$}
        \For{$i \in \{0, \ldots, N\}$}
        \State $q_1[i] \gets \Call{sin}{q[i]}$
        \EndFor
        \State $y \gets \Call{BLASGemv}{D, q_1}$
        \State \Return $y$
        \EndProcedure
      \end{algorithmic}
    \end{algorithm}

  \end{minipage}
  \hfill
  \begin{minipage}[t]{0.47\textwidth}
    \algrenewcommand{\alglinenumber}[1]{\color{magenta}\footnotesize#1}
    \begin{algorithm}[H]
      \caption{Functional batched einsum representation for {FEMOperator} in
      Algorithm~\ref{alg:intro_blas_overheads}.}
      \label{alg:functional_intro_blas_overheads}
      \begin{algorithmic}[1]
        \Procedure{FEMOperator}{$D$, $q$}
        \State $\lambda_A := \lambda i\;j.\;D[i,j]$
        \State $\lambda_x := \lambda j.\;\Call{sin}{q[j]}$
        \State $y \gets \Call{Einsum}{ij,j\to i, \lambda_A, \lambda_x}$
        \State \Return $y$
        \EndProcedure
      \end{algorithmic}
    \end{algorithm}
  \end{minipage}
\end{figure*}

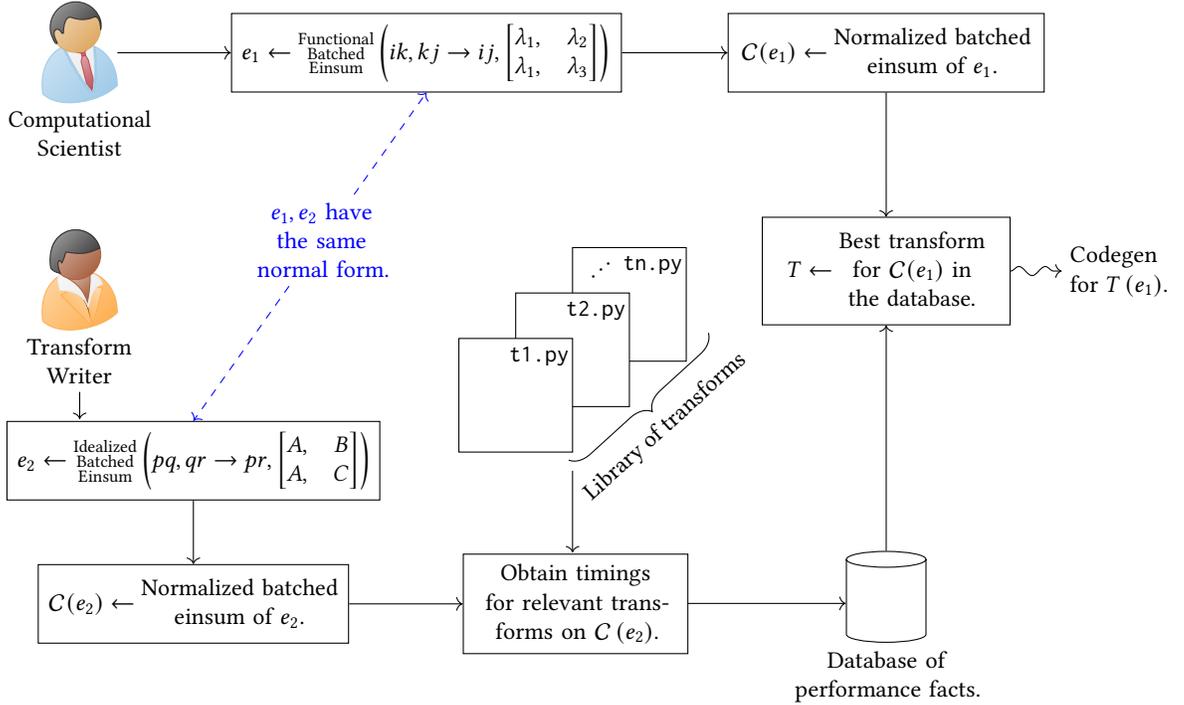
\begin{figure}
  \begin{tikzpicture}[
      x={(1cm,0cm)},
      y={(0cm,1cm)},
      z={(0.6cm,-0.4cm)}, %
    ]
    \def\W{\textwidth};
    \def\H{0.8*\textwidth};
    \node[dave,minimum size=1cm] (Dave) at (0.05*\W, 0.85*\H)  {};
    \node[yshift=-3ex, text width=14 ex, align=center] (DaveName) at
    (Dave.south)
    {Computational Scientist};
    \node[draw, rectangle, anchor=west, text width=.325*\W, inner sep
    = 1ex, align=center] (e1) at ($(Dave.east) + (0.1*\W, 0)$){$
      \begin{aligned}
        &e_1\leftarrow\operatorname{\substack{\text{Functional}\\\text{Batched}\\\text{Einsum}}}\left(ik,kj\rightarrow
          ij,
          \begin{bmatrix}
            \lambda_1,&\lambda_2\\
            \lambda_1,&\lambda_3
        \end{bmatrix}\right)
      \end{aligned}$%
    };
    \draw[->] (Dave.east) -- (e1.west);

    \node[draw, rectangle, anchor=west, inner sep = 1ex, text
    width=0.26*\W, align=center] (canon_e1) at ($(e1.east) + (0.093*\W, 0)$) {$
      \begin{aligned}
        \mathcal{C}(e_1) \leftarrow
        \begin{matrix}
          \text{Normalized batched}\\
          \text{einsum of } e_1.
        \end{matrix}
      \end{aligned}$
    };
    \node[draw, rectangle, anchor=south west, inner sep = 1ex, text
    width=0.2*\W, align=center] (transform_k) at (0.65*\W, 0.55*\H) {
      $
      \begin{aligned}
        T \leftarrow
        \begin{matrix}
          \text{Best transform} \\
          \text{for } \mathcal{C}(e_1) \text{ in}\\
          \text{the database.}
        \end{matrix}
      \end{aligned}$
    };

    \draw[->] (e1.east) -- ++(0.05*\W, 0) |- (canon_e1.west);
    \draw[->] (canon_e1.south) -- (transform_k.north);
    \node[draw, cylinder, shape border rotate=90, minimum height=0.1*\H,
    minimum width=0.07*\W] (db) at ($(transform_k.south) + (0, -0.306*\H)$) {};
    \node[rectangle, align=center, text width=20ex, anchor=north] at
    (db.south) {Database of performance facts.};
    \draw[->] (db.north) -- (transform_k.south);
    \draw[->, decorate, decoration={snake, amplitude=0.005*\H}]
    (transform_k.east) -- ++(0.045*\W, 0) node[right, text width=0.1\W]
    {Codegen for $T\left(e_1\right)$.};

    \node[alice,minimum size=1cm] (Alice) at (0.05*\W, 0.6*\H)  {};
    \node[yshift=-3ex, text width=14 ex, align=center] (AliceName) at
    (Alice.south)
    {Transform Writer};
    \node[draw, rectangle, anchor=north,
    text width=.31*\W, inner sep = 1ex, align=center] (e2) at
    ($(Alice.south) + (0.1*\W, -0.1*\H)$){
      $
      \begin{aligned}
        &e_2\leftarrow\operatorname{\substack{\text{Idealized}\\\text{Batched}\\\text{Einsum}}}\left(pq,qr\rightarrow
          pr,
          \begin{bmatrix}
            A,&B\\
            A,&C
        \end{bmatrix}\right)
      \end{aligned}$
    };
    \draw[->] (AliceName.south) -- ++(0, -0.03*\H);

    \node[draw, rectangle, anchor=north, inner sep = 1ex, text
    width=0.255*\W, align=center] (canon_e2) at ($(e2.south)+(0, -0.07*\H)$) {$
      \begin{aligned}
        \mathcal{C}(e_2) \leftarrow
        \begin{matrix}
          \text{Normalized batched}\\
          \text{einsum of } e_2.
        \end{matrix}
      \end{aligned}$
    };
    \draw[->] (e2.south) -- (canon_e2.north);
    \node[draw, rectangle, anchor=west, inner sep=1ex, align=center, text
    width=20ex] (RecordTime)
    at ($(canon_e2.east) + (0.1*\W, 0)$) {Obtain timings for relevant
    transforms on $\mathcal{C}\left(e_2\right)$.};
    \draw[->] (canon_e2.east) -- (RecordTime.west);
    \draw[->] (RecordTime.east) |- (db.west);

    \node[draw, rectangle, fill=white, anchor=south west, minimum
    width=0.1*\W, minimum height=0.1*\W] (File1) at ($(Alice.east) +
    (0.4*\W, -0.09*\H)$) {};
    \node[draw, rectangle, fill=white, anchor=south west, minimum
    width=0.1*\W, minimum height=0.1*\W] (File2) at ($(File1.south west)
    - (1*0.05*\W, 1*0.05*\H)$) {};
    \node[draw, rectangle, fill=white, anchor=south west, minimum
    width=0.1*\W, minimum height=0.1*\W] (File3) at ($(File1.south west)
    - (2*0.05*\W, 2*0.05*\H)$) {};
    \node at ($(File1.north east) + (-0.03*\W, -0.02*\H)$) {\tt tn.py};
    \node at ($(File2.north east) + (-0.03*\W, -0.02*\H)$) {\tt t2.py};
    \node at ($(File3.north east) + (-0.03*\W, -0.02*\H)$) {\tt t1.py};
    \node[rotate=41, align=center] at ($0.5*(File1.north) +
    0.5*(File2.north)$) {$\ldots$};
    \draw [decorate,decoration={brace,amplitude=2ex,raise=1ex}]
    ($(File1.south east) + (0.013*\W, 0.04*\H)$) -- ($(File3.south east)
    + (-0.01*\W, -0.0*\H)$)
    node[midway,rotate=42,yshift=-5ex]{Library of transforms};
    \draw[->] ($(File3.south) + (0.05*\W, -0.02*\H)$) -- ++(0, -0.09*\H);

    \coordinate (e2NorthRight) at ($(e2.north) + (0.*\W, 0)$);
    \node[color=blue, align=center, text width=13ex]  (SameCanonCommment)
    at ($0.45*(e2NorthRight) + 0.55*(e1.south)$) {$e_1, e_2$ have the
    same normal form.};
    \draw[dashed, blue, ->] (SameCanonCommment) -- (e1.south);
    \draw[dashed, blue, ->] (SameCanonCommment) -- (e2NorthRight);
  \end{tikzpicture}
  \caption{The \textsc{Feinsum} approach to achieve \emph{separation of
    concerns}. Computational scientist expresses his workload as a
    functional batched einsum and retrieves optimizations from the
    database. Transformation writer implements transformations for
    idealized batched einsums and records their performance in the
    database. The normal form is integral in allow the reuse of
    transformation knowledge from the performance engineer to the
  high-level programmer.}\label{fig:feinsum_overview}
  \Description{Demonstrates the end-to-end workflow involved in
  {\textsc{Feinsum}}.}
\end{figure}

We have implemented the computation of the normal form, matching of functional
batched einsums to idealized batched einsums, as well as a prototype database
retrieval for optimized variants in an open-source \textsc{Python} library,
called \textsc{Feinsum}\footnote{\url{https://pypi.org/project/feinsum}}.  In
\textsc{Feinsum}, we incorporate an \textsc{SQL} database which maps canonical
batched einsums to high-performant code-transformations on target hardware. The
database enables computational scientists to straightforwardly incorporate
learned transformation knowledge into their programs. Additionally, it provides
valuable data points for transform writers to evaluate advanced auto-tuning
approaches. In Figure~\ref{fig:feinsum_overview}, we see how the computation of
normal form is an integral requirement for \textsc{Feinsum}.

In this work, we present the following contributions:
\begin{itemize}
  \item A definition of a computational primitive representing an unordered
    collection of einsum operations. We refer to this primitive as a ``batched
    einsum''.  Subprograms that perform batched einsums allow for the modeling
    of operand reuse between the individual einsums, providing a performance
    engineer to develop tailored implementations that leverage such reuse.

  \item A canonicalization algorithm for batched einsum expressions. The
    algorithm resolves the inherent ambiguities in Einstein notation arising
    from the commutativity of multiplication and the possibility of index
    renaming.

  \item An extension of batched einsums, called \emph{functional batched
    einsums}, that considers $\lambda$-expressions as its operands. This
    facilitates the transfer of code optimizations known for idealized batched
    einsum instances to practical programs, especially relevant for devices with
    high saturation arithmetic intensity.

  \item An open-source software system that implements the matching of functional
    batched einsums to their idealized forms and the mapping of batched einsums
    to a canonical form to create and manage a database. This database maps the
    canonical form of some instances of batched einsums to their corresponding
    near-optimal code transformations.
\end{itemize}

\section{Preliminaries}
\subsection{Multidimensional Arrays}
We will use the following definition of a multidimensional array.

\begin{definition}
  A multidimensional array is a data type that has the following features:
  \begin{description}
    \item[\textsc{Dim}] Dimensionality of the array. If $a$ is a
      multidimensional array then $\operatorname{Dim}(a)$ evaluates to,
      a non-negative integer.
    \item[\textsc{Shape}] Shape of the array. If $a$ is a multidimensional
      such that $\operatorname{Dim}(a) = N$, then
      $\operatorname{Shape}(a)$ evaluates to a sequence of $N$
      non-negative integers.
    \item[\textsc{Dtype}] Data type of an element of the array. If $a$ is an
      array, then $\operatorname{Dtype}(a)$ evaluates to a numeric data
      type.  We refer
      the reader to {\normalfont\textsc{NumPy}~\cite{harris2020array}} for an
      exhaustive set of numeric-data types.
  \end{description}
  It further permits \emph{index access} operation: if $a$ is a multidimensional
  array with shape $(d_1, d_2, \ldots, d_n)$ and $(i_1, i_2, \ldots, i_n)$ is a
  sequence of non-negative integers such that $i_k < d_k$ for all $k \in \{1,
  \ldots n\}$, then $\operatorname{IndexInto}(a$, $(i_1, i_2, \ldots, i_n))$
  yields a scalar value of the type $\operatorname{Dtype}(a)$. We often use
  $a[i_1, i_2, \ldots, i_n]$ as a shorthand for the same indexing operation.
\end{definition}

We now define the domain and codomain associated with an array as follows.

\begin{definition}[Array Domain]
  Let $A$ be an $n$-dimensional array with shape $(s_1,\ldots,s_n)$.
  The \emph{domain} of $A$, denoted $\operatorname{dom}(A)$, is the
  Cartesian product
  \[
    \operatorname{dom}(A)
    = \prod_{i=1}^{n} \{0, 1,\ldots,s_i-1\},
  \]
  that is, the set of all valid index tuples for accessing elements of $A$.
\end{definition}

\begin{definition}[Array Codomain]
  Let $A$ be a multidimensional array. The \emph{codomain} of $A$, denoted by
  $\operatorname{codomain}(A)$, is the set of all scalar values of type
  $\operatorname{Dtype}(A)$.
\end{definition}

\subsection{Einstein Notation}
The popular Einstein summation convention~\cite{einstein1922basis} was developed
as a notation for contraction operation over tensors. In scientific software, a
slightly modified version of that definition is used for contractions over
multidimensional arrays. We now introduce a grammar that permits computational
representation of expressions in Einstein summation form.

\begin{definition}[Index]\label{defn:index}
  An \textit{index}, $i$ is an expression with the grammar as shown in
  Listing~\ref{lstng:grammar_for_index}.
\end{definition}

\lstset{style=pythonstyle}
\lstset{style=bnfstyle, frame=single}
\begin{lstlisting}[caption={Grammar for an index},
                   label={lstng:grammar_for_index}]
<index> = "a" | "b" | "c" | .. | "z"
\end{lstlisting}

We use the \textit{index} expression to define an index list.

\begin{definition}[Index List]
  An index list $\mathcal{I}$ is an expression built from a sequence of index
  expressions as shown in the grammar of Listing~\ref{lstng:grammar_for_index_set}.
\end{definition}

We use the subscript notation $\mathcal{I}_k$ to refer to the $k$-th index in
an index list.

\lstset{style=bnfstyle, frame=single}
\begin{lstlisting}[caption={Grammar for an index list, uses the \texttt{index} terminal
                             from Listing~\ref{lstng:grammar_for_index}},
                   label={lstng:grammar_for_index_set}]
<index_list> := <index>*
\end{lstlisting}

\begin{definition}[Length of an Index List]
  If $\mathcal{I} = (i_1, i_2, \ldots, i_n)$ is an index list, where
  $i_1, \ldots, i_n$ are index expressions, then the length of $\mathcal{I}$,
  denoted by $|\mathcal{I}|$, is equal to $n$.
\end{definition}

\begin{definition}[Einstein Summation (``einsum'')]
  An Einstein summation is a quadruple ($n$, $\mathcal{I}^{\mathrm{out}}$,
  $(\mathcal{A}^k)_{k=1}^{n}$, $(\mathcal{I}^{\mathrm{in}, k})_{k=1}^{n}$) that
  denotes an expression corresponding to a multidimensional array $R$, such that
  \begin{itemize}
    \item $n$ is a non-negative integer corresponding to number of inputs to
      the expression.
    \item $\mathcal{I}^{\mathrm{out}}$ is an index list, corresponding to the
      indexing pattern in $R$.
    \item $(\mathcal{A}_k)_{k=1}^{n}$ is a sequence of symbols for
      multidimensional arrays that form the inputs of the expression.
    \item $(\mathcal{I}^{\mathrm{in},k})_{k=1}^{n}$ is a sequence of index lists
      denoting the indexing behavior in the input multidimensional arrays.
    \item $R$ is evaluated as:
      \begin{align}\label{eq:define_einsum_output_r}
        R[\mathcal{I}^{\mathrm{out}}_1, \ldots,
        \mathcal{I}^{\mathrm{out}}_{|\mathcal{I}^{\mathrm{out}}|}]
        =
        \sum\limits_{\bigcup\limits_{j=1}^{n} (\mathcal{I}^{\mathrm{in},
        j} \setminus \mathcal{I}^{\mathrm{out}})}
        \left(
          \prod\limits_{k=1}^n \mathcal{A}^k[\mathcal{I}^{\mathrm{in},
            k}_1, \ldots, \mathcal{I}^{\mathrm{in},
          k}_{\operatorname{Dim}(\mathcal{A}^k)}]
        \right).
      \end{align}
  \end{itemize}
\end{definition}

It can be observed that for~\eqref{eq:define_einsum_output_r} to be sound,
the following properties must hold:
\begin{itemize}
  \item For all $k\in\{1,\ldots,n\}$, $|\mathcal{I}^{\mathrm{in},k}|$ must be
    equal to $\operatorname{Dim}\left(\mathcal{A}^k\right)$.
  \item $\operatorname{Dim}(R)$ must be equal to $|\mathcal{I}^{\mathrm{out}}|$.
  \item If $(k_1, i_1, k_2, i_2)$ satisfy
    $\mathcal{I}^{\mathrm{in}, k_1}_{i_1} = \mathcal{I}^{\mathrm{in},
    k_2}_{i_2}$, then $(\operatorname{Shape}(\mathcal{A}^{k_1}))_{i_1}$ must
    be equal to $(\operatorname{Shape}(\mathcal{A}^{k_2}))_{i_2}$.
  \item If $i \in \{1, \ldots, |\mathcal{I}^{\mathrm{out}}|\}$, then $
    \mathcal{I}^{\mathrm{out}}_i$ must be a subset of
    $\bigcup\limits_{k=1}^{n}\{\mathcal{I}^{\mathrm{in}, k}_1, \ldots,
    \mathcal{I}^{\mathrm{in},k}_{|\mathcal{I}^{\mathrm{in},k}|}\}$.
\end{itemize}

\begin{example}[Einsum]\label{example:einsum_intro}
  An einsum defined by the tuple $\left(2, \left(i, j\right), \left(\left(i,
    k\right), \left(k, j\right)\right), \left(A^{10 \times 4}, B^{4 \times
  10}\right)\right)$ corresponds to a rectangular matrix multiply operation,
  where the output $R^{10\times 10}$ is computed as:
  \[
    R\left[i, j\right] = \sum\limits_{k=0}^{3} A\left[i, k\right] \cdot
    B\left[k, j\right].
  \]
\end{example}

\begin{definition}[Equality of Index Lists]
  We say two index lists $\mathcal{I}$ and $\mathcal{I}'$ are equal, i.e.
  $\mathcal{I} =
  \mathcal{I}'$, iff:
  \begin{itemize}
    \item $|\mathcal{I}| = |\mathcal{I}'|$.
    \item For all $k\in\{1,\ldots,|\mathcal{I}|\}$, $\mathcal{I}_k =
      \mathcal{I}'_k$.
  \end{itemize}
\end{definition}

\begin{definition}[Equality of einsums]\label{defn:einsum_equalilty}
  We say two einsums $e_1$ defined by the tuple $(n$,
    $\mathcal{I}^{\mathrm{out}}$, $(\mathcal{I}^{\mathrm{in},k})_{k=1}^{n}$,
  $(\mathcal{A}^{k})_{k=1}^{n})$ and $e_2$ defined by the tuple $(n$,
    $\mathcal{I}'^{\mathrm{out}}$, $(\mathcal{I}'^{\mathrm{in},k})_{k=1}^{n}$,
  $(\mathcal{A}'^{k})_{k=1}^{n})$  to be equal i.e. $e_1 = e_2$,
  iff:
  \begin{itemize}
    \item $\mathcal{I}^{\mathrm{out}} = \mathcal{I}'^{\mathrm{out}}$.
    \item For all $k\in\{1, \ldots, n\}$, $\mathcal{I}^{\mathrm{in}, k} =
      \mathcal{I}'^{\mathrm{in},k}$.
    \item For all $k\in\{1, \ldots, n\}$, $\mathcal{A}^{k} =
      \mathcal{A}'^{k}$.
  \end{itemize}
\end{definition}

\begin{definition}[All indices in an einsum]
  If $e$ is an einsum which is defined by the tuple $(n$,
    $\mathcal{I}^{\mathrm{out}}$, $(\mathcal{I}^{\mathrm{in},k})_{k=1}^{n}$,
  $(\mathcal{A}^{k})_{k=1}^{n})$, then $\mathcal{I}^{\mathrm{all}}(e)$ denotes
  all the indices in the einsum's index notation, i.e.,
  \[
    \mathcal{I}^{\mathrm{all}}(e) =
    \bigcup\limits_{k=1}^{n}\mathcal{I}^{\mathrm{in}, k}.
  \]
\end{definition}

\section{Related Work}

Einsums are a fundamental $n$-d array operation seen across various domains of
scientific computing. This widespread occurrence of einsum makes it integral for
scientific computing frameworks to provide a primitive for it. Early
frameworks, like \textsc{BLAS}~\cite{lawson1979basic},
\textsc{LAPACK}~\cite{anderson1999lapack}, provided abstractions to implement
matrix-matrix multiplication and transposes which can be composed to perform
einsums.  {\AA}hlander et al.~\cite{aahlander2002einstein,aahlander2002software}
embraced this idea further and developed a C++-library that provided the einsum
primitive and used it in the context of high-level Partial Differential
Equations (PDEs) and iterative linear solvers.  Hirata et
al.~\cite{hirata2003tensor} proposed Tensor Contraction Engine (TCE), a
high-level $n$-d array manipulation framework that employed the Einstein
notation to describe quantum chemistry computations. TCE's index notation is
utilized by computational chemistry frameworks, like
NWChem~\cite{valiev2010nwchem}. \textsc{NumPy}~\cite{harris2020array} also
provides a primitive for Einstein summation and has since
become widely adopted by \textsc{Python}-based machine learning frameworks. As a
result, \textsc{TensorFlow}~\cite{abadi2015tensorflow},
\textsc{JAX}~\cite{frostig2018compiling},
\textsc{PyTorch}~\cite{paszke2019pytorch}, etc., provide the
``einsum''-primitive.  Vasilache et al.~\cite{vasilache2018tc} introduced Tensor
Comprehensions, a high-level DSL for expressing machine learning operations
using Einstein notation, which is subsequently lowered to machine code via the
polyhedral model. Rogozhnikov~\cite{rogozhnikov2022einops} developed an $n$-d
array library that relies on the index notation to specify array manipulation
primitives. Further, in Tensor Algebra Compiler, Kjolstad et
al.~\cite{kjolstad2017taco} extended the Einstein notation to be applicable over
sparse operands to map operations in Finite Element Methods (FEM), Sparse Neural
Networks, etc.

The ubiquity of einsum primitive across various computational frameworks has
resulted in a multitude of following work that optimize them for high
performance. Optimized BLAS implementations, such as
\textsc{OpenBLAS}~\cite{zhangOpenBLAS}, Intel \textsc{MKL}~\cite{Wang2014MKL},
\textsc{BLIS}~\cite{vanzeeBLIS}, \textsc{LibXSMM}~\cite{heineckelibxsmm},
provide optimized implementations of General Matrix-matrix multiplications
(GEMM). These high-performant GEMMs and transposes have been integral in early
tensor contraction algorithms, such as
Transpose-Transpose-GEMM-Transpose~\cite{hirata2003tensor} (TTGT).  Tensor
contractions generalize matrix multiplication to multidimensional arrays and
constitute a fundamental suboperation of einsums. They are typically paired with
contraction path optimization algorithms to evaluate einsums in a FLOP-optimal
manner. Lam et al.~\cite{lameinsumcomplexity1997} showed that this optimization
problem is NP-hard. Pfeifer et al.~\cite{pfeifer2014netcon} propose a pruning
strategy for the contraction path search space. Smith et
al.~\cite{Smith2018opteinsum} implement this search heuristic, along with
others, in \textsc{Opt-Einsum}, which has been subsequently used in
\textsc{NumPy}, \textsc{JAX}, etc. There has also been a wide array of work that
proposes domain-specific loop transformations for tensor contraction programs.
We refer the reader to the work by Springer et al.~\cite{bientinesi2018ttgt}
which includes a survey of tensor contraction approaches, such as TTGT, Loops
over GEMM, and nested loops approaches.  Additionally, the authors also propose
GEMM-like Tensor-Tensor multiplication (GETT) that avoids materializing the
transposes and performs transposition on the tiled operands which fit into the
cache, which end up being as operands of a GEMM micro-kernel. Kim et
al.~\cite{kim2019code} develop an algorithm for tensor contractions targeting
GPUs. They prefetch the tiled operands to scratchpad memory and construct the
tiled output via outer products.  The prefetching and outer product computation
is performed parallelly by the work-items of work-group. In TVM, Chen et
al.~\cite{chen2018tvm} propose an auto-tuning based lowering pass that explores
various loop transformations.  Their simulated annealing based auto-tuner is
shown to be competitive with simpler hand-tuned tensor contractions implemented
in \textsc{cuBLAS}.

The availability of high-performant tensor contractions / einsums has also led
to key applications that rely on these high-level routines.
Rockt{\"a}schel~\cite{rocktaschel2018einsum} shows this in the context of
machine-learning workloads. In a previous study~\cite{kulkarni2023domain}, we
demonstrated that einsums form the core operation in an end-to-end FEM
timestepper. Julien et al.~\cite{julien_toeinsumcompiler} demonstrate potential
profitability of decomposing array expressions into einsums. Barthels et
al.~\cite{barthels2021linnea} present a compiler that accepts array expressions
and rewrites them into optimized BLAS and LAPACK routines.

To exploit data reuse across multiple invocations of linear algebra primitives
and to improve utilization of wide SIMD architectures, batched linear algebra
operators have been studied. Abdelfattah et
al.~\cite{abdelfattah_2016_batchedgemms} developed an auto-tuning based strategy
for generating optimized batched GEMM routines on GPU architectures. Dongarra et
al.~\cite{dongarra_2017_batchedsolves} employed cross-matrix vectorization to
implement efficient batched GEMM and triangular solvers on wide SIMD
architectures. Kim et al.~\cite{kim_2018_batchedtcs} proposed a lowering
strategy for batches of tensor contractions.  In this work, we introduce a
computational primitive for a collection of einsum operations, referred to as
a batched einsum. Existing approaches to batching einsums typically introduce an
additional index to represent the batch dimension, however, such formulations
fail to capture fine-grained operand reuse and enforce arbitrary materialization
constraints that results in performance degradation.

To summarize, einsums comprise a key computational construct in scientific
computing codes. Many algorithms to compile einsums have been proposed, however
as evident from Table 3,4 of ~\cite{bientinesi2018ttgt}, no algorithm strictly
achieves roofline for all cases. Consequently, many scientific computing
frameworks choose to use handwritten kernels. We see this in the context of
state-of-the-art Finite Element
Solvers~\cite{klockner2009nodal,kronbichler2019fastfem}. To address these
issues, in this work, we propose a software framework that helps to tabulate near-roofline
kernels for instances of batched einsums, while allowing the tabulated knowledge
be reusable in a user-driven computational setting. A key requirement for
encoding an instance of batched einsum in a table for retrieval is the ability
to map it to a \emph{normal form}. To the best of our knowledge, no prior work
has addressed the problem of computing the normal form for einsum expressions,
let alone batched einsums.

\section{Batched Einstein Summation}\label{sec:define_batched_einsum}
It is common in computational science applications to invoke a kernel
corresponding to an einsum instance repeatedly, with substantial overlap in the
data accessed across invocations. Such patterns arise, for example, within the
\emph{hot loops} of FEM solvers, where expressions of the form
\[
  \sum_{j, x} J[x,r,e] D[x,i,j] u_{k}[e,j],\qquad{k\in\{1, \ldots, N_{\mathrm{field}}\}},
\]
occur frequently. Here, $J$ is the Jacobian, $D$ is the divergence matrix
associated with the reference cell and $u_{k}$ the field vectors defined over a
common discretization with affine geometry mapping. When these
$N_{\text{field}}$ einsums are evaluated independently, the shared operands $J$
and $D$ are repeatedly loaded, resulting in additional DRAM-to-processing unit
traffic and degraded performance.  Similar patterns arise in electronic
structure calculations as well as in linear and nonlinear solvers.

To capture this, we introduce an operation that encompasses multiple einsum
operations, referred to as the \textit{Batched Einstein Summation} operator.

\begin{definition}[Batched Einstein Summation]
  A Batched Einstein Summation is a quintuple $(b$, $n$,
    $\mathcal{I}^{\mathrm{out}}$, $(\mathcal{I}^{\mathrm{in},j})_{j=1}^{n}$,
  $((\mathcal{A}^{i,j})_{j=1}^{n})_{i=1}^{b})$, which evaluates a sequence of
  $b$ multidimensional arrays $(R_i)_{i=1}^b$. Each $R_i$ is the result of an
  einsum operation defined by the quadruple $(n$, $\mathcal{I}^{\mathrm{out}}$,
  $(\mathcal{I}^{\mathrm{in},j})_{j=1}^{n}$, $(\mathcal{A}^{i,j})_{j=1}^{n})$
  for $i \in \{1, 2, \ldots, b\}$.
\end{definition}

We will now provide an example of a \textit{batched einsum}.

\begin{example}[Batched einsum]
  A batched einsum defined by the tuple:
  \begin{equation*}
    \left(2, 3, \left(i, j\right), \left(\left(i,
      k\right), \left(k, l\right), \left(l, j\right)\right),
      \left(\left(A^{10 \times
        10}, B^{10 \times 10}, C^{10 \times 10}\right), \left(B^{10 \times
    10}, C^{10 \times 10}, D^{10 \times 10}\right)\right)\right)
  \end{equation*}
  corresponds to the
  computation of arrays $R_1^{10\times 10}$, and, $R_2^{10\times 10}$ that are
  evaluated as:
  \begin{align}
    \begin{split}
      R_1\left[i, j\right] &= \sum\limits_{k=0}^{9}\sum\limits_{l=0}^{9}
      A\left[i, k\right]
      B\left[k, l\right] C\left[l, j\right],\\
      R_2\left[i, j\right] &= \sum\limits_{k=0}^{9}\sum\limits_{l=0}^{9}
      B\left[i, k\right]
      C\left[k, l\right]  D\left[l, j\right].
    \end{split}
  \end{align}
\end{example}

We will now define certain operations on a batched einsum that will be useful in
subsequent sections.

\begin{definition}[Equality of batched einsums]
  Two batched einsums, $e_1 = \Bigl(b$, $n$,
    $\mathcal{I}^{\mathrm{out}}$,
    $\left(\mathcal{I}^{\mathrm{in},j}\right)_{j=1}^{n}$,
  $\left(\left(\mathcal{A}^{i,j}\right)_{j=1}^{n}\right)_{i=1}^b\Bigr)$, and,
  $e_2=\Bigl(b$, $n$, $\mathcal{I}'^{\mathrm{out}}$,
    $\left(\mathcal{I}'^{\mathrm{in},j}\right)_{j=1}^{n}$,
  $\left(\left(\mathcal{A}'^{i,j}\right)_{j=1}^{n}\right)_{i=1}^b\Bigr)$ are
  equal i.e. $e_1 = e_2$, iff for all $i\in\{1,\ldots,b\}$, the einsums defined
  by $\Bigl(n$, $\mathcal{I}^{\mathrm{out}}$,
    $(\mathcal{I}^{\mathrm{in},j})_{j=1}^{n}$,
  $\left(\mathcal{A}^{i,j}\right)_{j=1}^{n}\Bigr)$ is equal to the einsum
  defined by $\Bigl(n$, $\mathcal{I}'^{\mathrm{out}}$,
    $(\mathcal{I}'^{\mathrm{in},j})_{j=1}^{n}$,
  $\left(\mathcal{A}'^{i,j}\right)_{j=1}^{n}\Bigr)$.
\end{definition}

\begin{definition}[All arguments of a batched einsum]
  Let $e = \Bigl(b$, $n$, $\mathcal{I}^{\mathrm{out}}$,
    $\left(\mathcal{I}^{\mathrm{in},j}\right)_{j=1}^{n}$,
  $\left(\left(\mathcal{A}^{i,j}\right)_{j=1}^{n}\right)_{i=1}^{b}\Bigr)$ be a
  batched einsum, we obtain all arguments of the einsum i.e.
  $\mathcal{A}^{\mathrm{all}}(e)$ as
  \[
    \mathcal{A}^{\mathrm{all}}(e) =
    \bigcup\limits_{i=1}^{b}\{\mathcal{A}^{i,1}, \ldots, \mathcal{A}^{i,n}\}.
  \]
\end{definition}

\section{Isomorphic Batched Einstein
Summations}\label{sec:isomorphic_batched_einsum}
In Figure~\ref{fig:demo_sameish_einsums}, we highlighted the non-canonical
nature of batched einsum expressions, where syntactically different batched
einsum expressions can correspond to the same operation. In this section, we
provide a formal definition of equivalent classes of batched einsums, which we
refer to as \textit{isomorphic batched einsums}.

First, we introduce the concept of isomorphic einsums and then use
this concept to define isomorphic batched einsums.

\begin{definition}[Einsum Isomorphism]
  We say two einsums, $e_1 = \left(n,
    \mathcal{I}^{\mathrm{out}}, \left(\mathcal{I}^{\mathrm{in},j}\right)_{j=1}^{n},
  \left(\mathcal{A}^j\right)_{j=1}^{n}\right)$, and, $e_2 = \Bigl(n,
    \mathcal{I}'^{\mathrm{out}}$,
    $\left(\mathcal{I}'^{\mathrm{in},j}\right)_{j=1}^{n},
  \left(\mathcal{A}'^j\right)_{j=1}^{n}\Bigr)$,  to be isomorphic i.e.
  $e_1 \simeq e_2$,
  if $\Bigl|\mathcal{I}^{\mathrm{all}}(e_1)\Bigr|  =
  \Bigl|\mathcal{I}^{\mathrm{all}}(e_2)\Bigr|$ and there exists three substitution
  mappings $\sigma^j$, $\sigma^{\mathcal{I}}$, and,
  $\sigma^{\mathcal{A}}$  such that:
  \begin{itemize}
    \item $\sigma^{j}:
      \{1, 2, \ldots, n\} \to \{1, 2, \ldots, n\}$ is a bijective mapping
      corresponding to the operand ordering between the two einsums.

    \item $\sigma^{\mathcal{I}}:
      \mathcal{I}^{\mathrm{all}}(e_2)\to\mathcal{I}^{\mathrm{all}}(e_1)$ is
      a bijective mapping
      corresponding to the index name mapping between the two einsums.

    \item $\sigma^{\mathcal{A}}: \{A'^1, \ldots, A'^n\}\to\{A^1, \ldots,
      A^n\}$ is a bijective mapping corresponding to the argument name mapping
      between the two einsums.

    \item For all $j\in\{1, 2, \ldots, n\}$,
      $\operatorname{Shape}\left(\mathcal{A}^{j}\right) =
      \operatorname{Shape}\left(\sigma^{\mathcal{A}}\left(\mathcal{A}'^{\sigma^j\left(j\right)}\right)\right)$
      and
      $\operatorname{Dtype}\left(\mathcal{A}^{j}\right) =
      \operatorname{Dtype}\left(\sigma^{\mathcal{A}}\left(\mathcal{A}'^{\sigma^j\left(j\right)}\right)\right)$.

    \item For all $j\in\{1, 2, \ldots, n\}$ and $d\in\left\{1, 2,\ldots,
      |\mathcal{I}^{d, in}|\right\}$, $|\mathcal{I}^{\mathrm{in},j}| =
      |\mathcal{I}'^{\mathrm{in},\sigma^{j}(j)}|$ and
      $\mathcal{I}^{\mathrm{in},j}_d =
      \sigma^{\mathcal{I}}\left(\mathcal{I}'^{\mathrm{in},\sigma^j\left(j\right)}_d\right)$.

    \item $\Bigl|\mathcal{I}^{\mathrm{out}}\Bigr| =
      \Bigl|\mathcal{I}'^{\mathrm{out}}\Bigr|$.

    \item For all $d \in \left\{1, 2, \ldots, |\mathcal{I}^{\mathrm{out}}|\right\}$,
      $\mathcal{I}^{\mathrm{out}}_d =
      \sigma^{\mathcal{I}}\left(\mathcal{I}'^{\mathrm{out}}_d\right)$.
  \end{itemize}
\end{definition}

\begin{example}[Isomorphic einsums]
  The einsums $e_1$ defined by the tuple $\Bigl(2$, $\left(i, j\right)$,
    $\left(\left(i, k\right), \left(k, j\right)\right)$, $\left(A^{10\times
  4}, B^{4\times 10}\right)\Bigr)$ and $e_2$ defined by the tuple
  $\Bigl(2$, $\left(p, q\right)$,
    $\left(\left(r, q\right), \left(p, r\right)\right)$,
    $\left(X^{4\times 10}, Y^{10\times
  4}\right)\Bigr)$ are isomorphic with the substitution mappings:
  \begin{itemize}
    \item $\sigma^j = \{1\mapsto 2; 2\mapsto 1\}$.
    \item $\sigma^{\mathcal{I}} = \{p\mapsto i; q\mapsto j; r\mapsto k\}$.
    \item $\sigma^{\mathcal{A}} = \{Y\mapsto A; X\mapsto B\}$.
  \end{itemize}
\end{example}

\begin{definition}[Batched einsum Isomorphism]\label{defn:isomorphic_batched_einsum}
  We say two batched einsums, $e_1 = \Bigl(b$, $n$,
    $\mathcal{I}^{\mathrm{out}}$,
    $\left(\mathcal{I}^{\mathrm{in},j}\right)_{j=1}^{n}$,
  $\left(\left(\mathcal{A}^{i, j}\right)_{j=1}^{n}\right)_{i=1}^b\Bigr)$, and
  $e_2 = \Bigl(b$, $n$, $\mathcal{I}'^{\mathrm{out}}$,
    $\left(\mathcal{I}'^{\mathrm{in},j}\right)_{j=1}^{n}$,
    $\left(\left(\mathcal{A}'^{i,
  j}\right)_{j=1}^{n}\right)_{i=1}^b\Bigr)$, to be isomorphic i.e. $e_1 \simeq
  e_2$, if there exists four substitution
  mappings $\sigma^i$, $\sigma^j$, $\sigma^{\mathcal{I}}$ and,
  $\sigma^{\mathcal{A}}$ such that:
  \begin{itemize}
    \item $\sigma^i: \{1, 2, \ldots, b\} \to \{1, 2, \ldots, b\}$ is a
      bijective mapping.
    \item $\sigma^{\mathcal{A}}: \mathcal{A}^{\mathrm{all}}(e_2) \to
      \mathcal{A}^{\mathrm{all}}(e_1)$ is a bijective mapping.
    \item For all $i \in \{1, 2, \ldots, b\}$, the einsums defined by $\Bigl(n$,
        $\mathcal{I}^{\mathrm{out}}$, $(\mathcal{I}^{\mathrm{in},j})_{j=1}^{n}$,
      $\left(\mathcal{A}^{i, j}\right)_{j=1}^{n}\Bigr)$ is isomorphic to
      $\Bigl(n$,
        $\mathcal{I}'^{\mathrm{out}}$, $(\mathcal{I}'^{\mathrm{in},j})_{j=1}^{n}$,
        $\left(\mathcal{A}'^{\sigma^{i}\left(i\right),
      j}\right)_{j=1}^{n}\Bigr)$ using the
      substitution mappings $\sigma^{j}$, $\sigma^{\mathcal{I}}$ and
      $\sigma_i^{\mathcal{A}}$, where $\sigma_i^{\mathcal{A}} =
      $\\ $\left\{\mathcal{A}^{i,j}\mapsto\sigma^{\mathcal{A}}\left(\mathcal{A}'^{\sigma^{i}\left(i\right),\sigma^{j}\left(j\right)}\right):
      j\in\{1, \ldots, n\}\right\}$.
  \end{itemize}
\end{definition}

\begin{example}[Isomorphic batched einsums]
  Consider the batched einsums $e_1$ defined by
  \begin{equation*}
    \left(2, 4,
      \left(i\right),
      \left(\left(i,j,k\right), \left(i,k\right), \left(i,j\right),
      \left(i,j\right)\right),
      \left(\left(A^{5\times10\times10}, B^{5\times10},
        C^{5\times10}, D^{5\times10}\right),\left(A^{5\times10\times10},
    B^{5\times10}, C^{5\times10}, B^{5\times10}\right)\right)\right),
  \end{equation*}
  and $e_2$ defined by
  \begin{equation*}
    \left(2,
      4,
      \left(i\right),
      \left(\left(i,k,j\right), \left(i,k\right), \left(i,k\right),
      \left(i,j\right)\right),
      \left(\left(P^{5\times10\times10}, S^{5\times10},
        R^{5\times10}, S^{5\times10}\right), \left(P^{5\times10\times10},
      Q^{5\times10}, R^{5\times10}, S^{5\times10}\right)\right)
    \right),
  \end{equation*}
  then $e_1\simeq e_2$, using the substitution mappings:

  \begin{itemize}
    \item $\sigma^i = \{1\mapsto 2; 2\mapsto 1\}$.
    \item $\sigma^j = \{1\mapsto 1; 2\mapsto 4; 3\mapsto 3; 4\mapsto 2\}$.
    \item $\sigma^{\mathcal{I}} = \{i\mapsto i; j\mapsto k; k\mapsto j\}$.
    \item $\sigma^{\mathcal{A}} = \{P\mapsto A; S\mapsto B; R\mapsto C;
      Q\mapsto D\}$.
  \end{itemize}
\end{example}

\section{A Canonical form for Batched
Einsums}\label{sec:canonicalization_via_directed}
A direct implication of two einsums, denoted $e_1$ and $e_2$, being isomorphic
is that an algorithmic procedure developed for evaluating $e_1$ can likewise be
used to evaluate $e_2$, provided that the corresponding variables are suitably
relabeled. One approach to determine if two einsums are isomorphic is by
defining a \textit{canonical form} for batched einsums such that isomorphic
batched einsums have equal canonicalized representations.

\begin{definition}[Canonicalization of a batched einsum]
  \label{defn:batched_einsum_canonical_form}
  Let $B$ denote the set of all batched einsums. A function $\mathcal{C}: B \to
  B$ is called a \emph{canonicalization function} if, for any batched einsum
  $e$, its canonical form $\mathcal{C}(e)$ satisfies
  \[
    e \simeq e' \;\Leftrightarrow\; \mathcal{C}(e) = \mathcal{C}(e'),
  \]
  for all batched einsums $e' \in B$.
\end{definition}
In this section, we present one such \textit{canonicalization function}.  The
procedure consists of three steps: (i) mapping a batched einsum to a colored
directed graph, called the \emph{induced graph}, (ii) canonicalizing the induced
graph, and (iii) mapping the canonicalized graph back to a batched einsum. The
mappings between batched einsums and induced graphs are detailed in
Appendix~\ref{sec:to_directed_induced_graph} and
Appendix~\ref{sec:from_directed_induced_graph}. We will now employ these
mappings to develop a canonicalization algorithm based on colored graph
canonicalization~\cite{mckay1981practical}.

We describe our batched einsum canonicalization algorithm in the procedure
\textsc{CanonicalizeBatchedEinsum} of
Algorithm~\ref{alg:canonicalize_batched_einsum}.  The procedure computes
$e' = \mathcal{C}(e)$, together with the substitution mappings
$\sigma^{\mathcal{A}}$ and $\sigma^{\mathcal{I}}$. The mappings
$\sigma^{\mathcal{A}}$ and $\sigma^{\mathcal{I}}$ map the array and index
symbols of $e'$ to their corresponding symbols in $e$, as specified in
Definition~\ref{defn:isomorphic_batched_einsum}.

Algorithm~\ref{alg:canonicalize_batched_einsum} begins by constructing an
induced graph representing the batched einsum $e$
(line~\ref{algline:call_to_induced_dag}). The components $A$, $c$,
$\iota_{\mathrm{dtype}}$, $\iota_{\mathrm{length}}$, $\iota_{\mathrm{index}}$,
and $\iota_{\mathrm{arg}}$ correspond to the components of an induced graph as
defined in Definition~\ref{defn:induced_directed_graph}. $A$ is the adjacency
matrix corresponding to the edges in the directed graph, $c$ is a column vector
corresponding to the colors of the nodes of the graph, $\iota_{\mathrm{dtype}}$
is the mapping from labels of nodes corresponding to data-types in the induced
graph to their numeric data types, $\iota_{\mathrm{length}}$ is the mapping from
labels of nodes corresponding to axis lengths in the induced graph to their axis
lengths, and,  $\iota_{\mathrm{index}}$ is the mapping from labels of nodes
corresponding to the indices of $e$ to their indices.

We then invoke a graph canonicalizer to generate a relabeling map that
associates the nodes of the constructed graph with their counterparts in the
canonical graph.
Lines~\ref{algline:start_build_adj_graph}--\ref{algline:end_build_adj_graph}
apply this relabeling map to produce the canonical form of the induced graph. At
line~\ref{algline:call_to_batched_einsum}, we employ the batched einsum
reconstruction procedure from Appendix~\ref{sec:from_directed_induced_graph} to
obtain $e'$, the reconstructed batched einsum. As part of this reconstruction,
$\iota_{\mathrm{index}}^{\mathrm{inferred}}$ and
$\iota_{\mathrm{arg}}^{\mathrm{inferred}}$ denote the enumerations of indices
and arguments selected during the process, as detailed
in~\ref{step:reconstruct_digraph_iota_index}
and~\ref{step:reconstruct_digraph_iota_arg} of
Appendix~\ref{sec:from_directed_induced_graph}. Finally, in
lines~\ref{algline:start_build_subst_map}--\ref{algline:end_build_subst_map}, we
construct the argument and index substitution mappings, denoted by
$\sigma^{\mathcal{A}}$ and $\sigma^{\mathcal{I}}$, respectively.

\begin{algorithm}[H]
  \caption{Canonicalizing a batched einsum}\label{alg:canonicalize_batched_einsum}
  \begin{algorithmic}[1]
    \Require Procedures \textsc{IndexName}, \textsc{ArgName}. These procedures
    accept a non-negative integer and deterministically return a unique index
    name and a unique array operand name for a batched einsum respectively. See
    Appendix~\ref{sec:from_directed_induced_graph} for formal definitions.

    \Procedure{CanonicalizeBatchedEinsum}{$e$}
    \State $A, c, \iota_{\mathrm{dtype}}, \iota_{\mathrm{length}},
    \iota_{\mathrm{index}}, \iota_{\mathrm{arg}} \gets$
    \Call{ToInducedGraph}{$e$}\Comment{See
    Appendix~\ref{sec:to_directed_induced_graph}.}\label{algline:call_to_induced_dag}

    \State RelabelingMap $\gets$ \Call{GetCanonLabel}{$A$, $c$} \Comment{Any
      graph canonization~\cite{junttila_2011_bliss, mckay2014practical}
    implementation.}

    \State $N \gets$ \# nodes in the graph corresponding to $A$.
    \State $A' \gets$ matrix of zeros of size $N \times N$.
    \State $c' \gets$ vector of zeros of size $N$.
    \State $\iota'_{\mathrm{dtype}},\iota'_{\mathrm{length}} \gets$ empty map.

    \For{$i \in \{1, \ldots, N\}$}\label{algline:start_build_adj_graph}
    \For{$j \in \{1, \ldots, N\}$}
    \State $A'[\mathrm{RelabelingMap}[i], \mathrm{RelabelingMap}[j]] \gets A[i, j]$
    \EndFor
    \State $c'[\mathrm{RelabelingMap}[i]] \gets c[i]$
    \If{$i \in \iota_{dtype}$}
    \State $\iota'_{\mathrm{dtype}}[\mathrm{RelabelingMap}[i]] \gets
    \iota_{\mathrm{dtype}}[i]$
    \EndIf
    \If{$i \in \iota_{length}$}
    \State $\iota'_{\mathrm{length}}[\mathrm{RelabelingMap}[i]] \gets
    \iota_{\mathrm{length}}[i]$
    \EndIf
    \EndFor\label{algline:end_build_adj_graph}

    \State $e', \iota_{\mathrm{index}}^{\mathrm{inferred}},
    \iota_{\mathrm{arg}}^{\mathrm{inferred}} \gets$
    \Call{ToBatchedEinsum}{$A'$, $c'$, $\iota_{\mathrm{dtype}}'$,
    $\iota_{\mathrm{length}}'$, \textsc{IndexName}, \textsc{ArgName}}\Comment{See
    Appendix~\ref{sec:from_directed_induced_graph}.}\label{algline:call_to_batched_einsum}

    \State
    \State $\sigma^{\mathcal{A}}, \sigma^{\mathcal{I}} \gets $ empty maps.

    \algstore{canonicalizepseudocode}
  \end{algorithmic}
\end{algorithm}

\begin{algorithm}[H]
  \begin{algorithmic}[1]
  \algrestore{canonicalizepseudocode}

    \For{$i \in \{1, \ldots, N\}$}\label{algline:start_build_subst_map}
    \If{$i \in \iota_{\mathrm{index}}$}
    \State oldName $\gets \iota_{\mathrm{index}}[i]$
    \State newName $\gets
    \operatorname{IndexName}\left(\iota_{\mathrm{index}}^{\mathrm{inferred}}[\mathrm{RelabelingMap}[i]]\right)$
    \State $\sigma^{\mathcal{I}}[\mathrm{newName}] \gets \mathrm{oldName}$
    \EndIf
    \If{$i \in \iota_{\mathrm{arg}}$}
    \State oldName $\gets \iota_{\mathrm{arg}}[i]$
    \State newName $\gets
    \operatorname{ArgName}\left(\iota_{\mathrm{arg}}^{\mathrm{inferred}}[\mathrm{RelabelingMap}[i]]\right)$
    \State $\sigma^{\mathcal{A}}[\mathrm{newName}] \gets \mathrm{oldName}$
    \EndIf
    \EndFor\label{algline:end_build_subst_map}

    \State \Return $\left(e', \sigma^{\mathcal{A}}, \sigma^{\mathcal{I}}\right)$
    \EndProcedure
  \end{algorithmic}
\end{algorithm}

\begin{remark}
  Since our canonicalization algorithm incorporates graph canonicalization, it
  satisfies the idempotence criterion, i.e.  for all batched einsums $e$,
  $\mathcal{C}\left(e\right) = \mathcal{C}\left(\mathcal{C}\left(e\right)\right)$.
\end{remark}

\begin{example}
  For the einsum $e_1$ defined by $\Bigl(1$, $2$, $\left(i\right)$,
    $\left(\left(i j\right),
  \left(i k\right)\right)$, $\left(\left(A, B\right)\right)\Bigr)$, with
  $\operatorname{Shape}(A)=\operatorname{Shape}(B) = (72, 18)$ and
  $\operatorname{Dtype}(A)=\operatorname{Dtype}(B) = \mathrm{F64}$, we get
  $\mathcal{C}(e_1)$ as the batched einsum defined by $\Bigl(1$, $2$,
    $\left(a\right)$, $\left(\left(a b\right),
  \left(a c\right)\right)$, $\left(\left(A_0, A_1\right)\right)\Bigr)$, with
  $\operatorname{Shape}(A_0)=\operatorname{Shape}(A_1) = (72, 18)$ and
  $\operatorname{Dtype}(A_0)=\operatorname{Dtype}(A_1) = \mathrm{F64}$.

  For the einsum $e_2$ defined by $\Bigl(1$, $2$, $\left(i\right)$,
    $\left(\left(i k\right), \left(i j\right)\right)$, $\left(\left(X,
  Y\right)\right)\Bigr)$, with
  $\operatorname{Shape}(X)=\operatorname{Shape}(Y) = (72, 18)$
  and $\operatorname{Dtype}(X)=\operatorname{Dtype}(Y) = \mathrm{F64}$, we get
  $\mathcal{C}(e_2)$ as the batched einsum defined by $\Bigl(1$, $2$,
    $\left(a\right)$, $\left(\left(a b\right), \left(a c\right)\right)$,
  $\left(\left(A_0, A_1\right)\right)\Bigr)$, with
  $\operatorname{Shape}(A_0)=\operatorname{Shape}(A_1) = (72, 18)$ and
  $\operatorname{Dtype}(A_0)=\operatorname{Dtype}(A_1) = \mathrm{F64}$. Since
  $\mathcal{C}(e_1) = \mathcal{C}(e_2)$, $e_1$ must be isomorphic to $e_2$.
\end{example}

It is worth noting that multiple canonicalization functions for batched einsums
are possible. Our proposed approach relies on a graph canonicalization
procedure. Furthermore, replacing the underlying graph canonicalizer in our
approach with a different implementation would yield a different
canonicalization function.

This canonicalization will be instrumental in tabulating batched einsums in our
software system described in Section~\ref{sec:feinsum_abstraction}.

\section{Integration of Optimized Batched Einsums into Practical
Programs}\label{sec:feinsum_abstraction}
In this section, we introduce our strategy for integration and practical reuse
of optimized implementations of batched einsums within high-level programs.  We
have implemented a software system that can maintain a library of transformation
plans that operate on functional batched einsums.  These transformation plans
are added to the library either by a performance engineer or by an autotuner.
Additionally, our system maintains a database of empirically populated
performance facts of the transformations in the library for specialized
instances of functional batched einsums, where every function for an operand is
of the form $(i_1, \ldots, i_n) \mapsto A[i_1, \ldots, i_n]$, where $A$ is a
row-major ordered multidimensional array. We refer to such instances of
functional batched einsums as \emph{idealized functional batched einsums}. A
computational scientist uses these empirical facts known for idealized functional
batched einsums to optimize their program that is expressed as a functional
batched einsum. We have implemented this as a \textsc{Python} library called
\textsc{Feinsum}\footnote{\textsc{Feinsum} is open-source and publicly available
at \url{https://pypi.org/project/feinsum}.}.  We summarize the core components
of \textsc{Feinsum} and the manner in which the computational scientist and
transform writer interact with it in Figure~\ref{fig:feinsum_overview}.

A key design choice for software systems such as \textsc{Feinsum}, which adopt a
tabulation-based approach to provide optimized subprograms, is the structure of
the computational primitive.  Software systems such as \textsc{BLAS},
\textsc{oneDNN}~\cite{onednnintel2024}, and, \textsc{LAPACK}, provide the
computational scientist binaries containing optimized subroutines for the
computational primitives. These approaches accept operands as in-memory
buffers. However, this approach of accepting buffers for operands introduces
additional overheads associated with buffer allocation and initialization. We
have previously observed one such instance of overheads induced by abstraction in
Algorithm~\ref{alg:intro_blas_overheads}. To avoid such materialization-related
overheads we extend the concept of a batched einsum such that it allows lambda expressions
instead of materialized buffers as inputs. We call this a \emph{functional
batched einsum expression}. The library of transformations provided along
\textsc{Feinsum} operate on such functional versions of batched einsums. The
database of performance facts contains measurements for an idealized version of
functional batched einsums.  We will now formally define functional batched
einsums and the associated idealized functional batched einsum.

\begin{definition}[Functional Batched Einstein
  Summation]\label{defn:functional_batched_einsum}
  Let $e = (b, n, \mathcal{I}^{\mathrm{out}},
    (\mathcal{I}^{\mathrm{in},j})_{j=1}^{n},
  ((\mathcal{A}^{i,j})_{j=1}^{n})_{i=1}^{b})$ be a batched einsum.
  Let $\mathcal{M}$ be a mapping assigning to each
  $A \in \mathcal{A}^{\mathrm{all}}(e)$ a function
  \[
    \mathcal{M}(A) : \operatorname{dom}(A) \to \operatorname{codomain}(A).
  \]
  The \emph{functional batched einsum} defined by the pair $(e,\mathcal{M})$
  computes a sequence of $b$ multidimensional arrays $(R_i)_{i=1}^{b}$ such that,
  for each $i \in \{1,\ldots,b\}$,
  \[
    R_i[\mathcal{I}^{\mathrm{out}}_1,\ldots,\mathcal{I}^{\mathrm{out}}_n]
    =
    \sum_{\bigcup_{j=1}^{n}(\mathcal{I}^{\mathrm{in},j} \setminus
    \mathcal{I}^{\mathrm{out}})}
    \left(
      \prod_{k=1}^{n}
      \mathcal{M}(\mathcal{A}^{i,k})
      \bigl(
        \mathcal{I}^{\mathrm{in},k}_1,\ldots,
        \mathcal{I}^{\mathrm{in},k}_{\operatorname{Dim}(\mathcal{A}^k)}
      \bigr)
    \right).
  \]
\end{definition}

\begin{definition}[Idealized Functional Batched Einsum]
  Let $f = (e, \mathcal{M})$ be a functional batched einsum such that
  \[
    \mathcal{M}
    =
    \left\{
      A \mapsto
      \lambda i_1\;\ldots\;i_n\;.\;
      P_A[i_1,\ldots,i_n]
      \;:\;
      A \in \mathcal{A}^{\mathrm{all}}(e),\;
      n = \operatorname{Dim}(A)
    \right\},
  \]
  where $P_A$ denotes the underlying buffer of a row-major ordered
  multidimensional array associated with $A$ that is not aliased by any other
  array in the program. Then $f$ is called an \emph{idealized functional batched
  einsum}.
\end{definition}

At its core, \textsc{Feinsum} permits the computational scientist (or a
high-level compiler) to reuse high-performant transformations known for idealized
functional batched einsums to their programs resembling batched einsums. In
general, the profitability analysis for the functional batched einsum is
sensitive to the expressions in $\mathcal{M}$. Nevertheless, we argue there is a
broad class of choices for $\mathcal{M}$ that appear in real-world scenarios
where transfer of such profitability analysis applies. To make this concrete, we
compare the saturation arithmetic intensities of several microarchitectures with
the arithmetic intensities of the einsums in the TCCG benchmark
suite~\cite{tccg2016a}. The results of this comparison are shown in
Figure~\ref{fig:tccg_ai}. We observe that, on average, 60.4\% of the benchmark
entries are memory bound across GPUs such as the NVIDIA P100, NVIDIA Titan~V,
NVIDIA H100, and AMD MI250X. A consequence of an idealized functional batched
einsum being memory bound is that any functional batched einsum that performs
additional computations on the operands without altering the memory access
pattern exhibits similar performance characteristics.  Thereby, reusing
transformations developed for idealized functional batched einsums is often a
suitable and effective strategy.

\pgfplotstableread[col sep=space]{./tccg_ai.dat}\tccgaidatatable
\begin{figure}[h]
  \centering
  \begin{tikzpicture}
    \begin{semilogyaxis}[
        xtick=data,
        xticklabels from table={\tccgaidatatable}{xticklabel},
        xticklabel style={rotate=90},
        ylabel={\shortstack{FP64 Arithmetic Intensity \\ (in
        $\frac{\mathrm{\# FLOPS}}{\mathrm{\# Bytes}}$)}},
        xticklabel style = {font=\tiny},
        ylabel style = {font=\small},
        xmin=0.5,
        xmax=48.5,
        width=1\linewidth,
        height=0.4\linewidth,
        legend pos= north east,
        legend style= {font=\small}
      ]
      \addplot+[thick, forget plot] table[x=x, y=y]{\tccgaidatatable};
      \addplot[magenta, thick, densely dotted] coordinates {(0.5,14.9) (48.5,14.9)};
      \addlegendentry{AMD MI250X}
      \addplot[cyan, thick, dashdotted] coordinates {(0.5,12.55) (48.5,12.55)};
      \addlegendentry{Nvidia H100}
      \addplot[thick, dotted] coordinates {(0.5,9.41) (48.5,9.41)};
      \addlegendentry{Nvidia Titan V}
      \addplot[red, thick, dashed] coordinates {(0.5,7.24) (48.5,7.24)};
      \addlegendentry{Nvidia P100}
    \end{semilogyaxis}
  \end{tikzpicture}
  \caption{
    Arithmetic intensities (AI) for the TCCG benchmark suite with
    double-precision operands. The horizontal dashed lines indicate saturation
    AIs of modern accelerators, above which the workloads become compute-bound.
  }\label{fig:tccg_ai}
\end{figure}
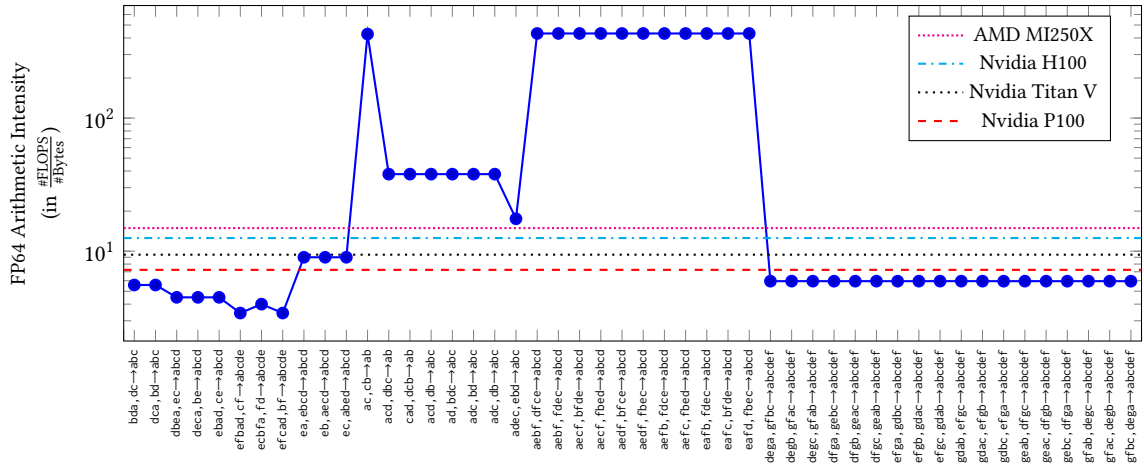

In our implementation of \textsc{Feinsum}, we utilize the \textsc{Loopy}
toolkit~\cite{klockner2014loo}. \textsc{Loopy} serves as both a language for
specifying the grammar and a framework for defining the transformations on
subprograms. We provide an overview of the Loopy Intermediate Representation
(IR) in Appendix~\ref{sec:preliminaries_loopy}. It is important to emphasize
that none of the concepts introduced in this work are dependent on
\textsc{Loopy}. The proposed approach can be readily implemented in any
high-level IR, such as \textsc{MLIR}~\cite{lattner_mlir_2021},
\textsc{Triton}~\cite{tillet_triton_2019} and
\textsc{Halide}~\cite{kelley_halide_2013}, provided that the representation
supports abstract syntax tree (AST) nodes for expressing multidimensional
arrays, loop nests and functions.

We now describe the flow shown in Figure~\ref{fig:feinsum_overview} in detail
using a sequence of examples. Listing~\ref{lstng:demo_feinsum_transform}
presents an example transformation from the transformation library.
Listing~\ref{lstng:demo_feinsum_subprogram} illustrates a retrieval from
\textsc{Feinsum}'s database that transforms an instance of functional batched
einsum.  Finally, Listing~\ref{lstng:demo_feinsum_opencl_subprogram} shows the
code generated for the retrieval in Listing~\ref{lstng:demo_feinsum_subprogram},
assuming that the transformation in Listing~\ref{lstng:demo_feinsum_transform}
is the best known transformation in the database. The generated code applies the
library-specified transformations after computing the required substitution
mappings. Moreover, the function evaluations corresponding to the operands are
fused into the kernel's inner loop.

\lstset{style=pythonstyle, frame=single,numbers=left,stepnumber=1}
\begin{lstlisting}[caption={Transformation writer's view of \textsc{Feinsum}.
  The transformation writer would implement a transformation plan targeting the
  kernel \texttt{K}.  Here, a transformation implemented for a batched einsum
  corresponding to two matrix-vector multiplications as represented by
  \texttt{ref\_ensm}.},
  label={lstng:demo_feinsum_transform}]
import feinsum as fs
import loopy as lp

A = fs.array("A", (96, 4))
B, C = fs.array("B", (4,)), fs.array("C", (4,))
ref_ensm = fs.batched_einsum("ij,j->i", [[A, B], [A, C]])

def transform(K):
    sigma = fs.identify_as_einsum(K, ref_ensm)
    # Tile i-loop in chunks of 32.
    K = lp.split_iname(K, sigma["i"], 32)
    # Precompute "A" since it is used twice.
    K = lp.precompute(K, sigma["A"], sweep_inames=())
    return K

if __name__ == "__main__":
    fs.record_facts(ref_ensm, device, __file__)
\end{lstlisting}

We now examine Listing~\ref{lstng:demo_feinsum_transform} in detail. The code
transformation is specified in the \texttt{transform} routine, which takes as
input an intermediate representation (IR) of a kernel and returns a transformed
kernel. The transformation is defined for the batched einsum $\left(2, 2, (i),
((ij), (j)), ((A, B), (A, C))\right)$. It begins by invoking
\texttt{identify\_as\_einsum}, which interprets the statements in the input
kernel as a functional batched einsum and computes a renaming map from the
variables of the reference einsum to those in the kernel IR. Interpreting
lower-level IR as a higher-level operation, such as a functional batched einsum,
is commonly referred to as \emph{raising}~\cite{chelini_2021_raisingmlir}.
Canonicalization is an essential component of \texttt{identify\_as\_einsum};
pseudocode for this procedure is provided in
Algorithm~\ref{alg:batched_ref_einsum_matching}. The resulting renaming map is
then used to apply a tiling transformation (\texttt{lp.split\_iname}) and to
precompute the $A$ operand in private memory, thereby avoiding redundant
accesses to global memory. The final line of the script invokes
\texttt{record\_facts}, which records the runtime of the idealized functional
batched einsum corresponding to \texttt{ref\_ensm} after applying the
\texttt{transform} routine. Before recording wall-clock times, we execute
several warm-up runs on the device to mitigate transient effects. The canonical
form of \texttt{ref\_ensm} is used as the identifier under which these
performance facts are stored.

\begin{algorithm}
  \caption{Algorithm to identify the kernel, $K$, as the reference
  einsum, $e_{\text{ref}}$.}\label{alg:batched_ref_einsum_matching}
  \begin{algorithmic}[1]
    \Procedure{IdentityAsEinsum}{$K$, $e_{\mathrm{ref}}$}
    \State $e_{\mathrm{raised}}, \sigma^{\mathcal{A}}_{\mathrm{raised}},
    \sigma^{\mathcal{I}}_{\mathrm{raised}} \gets$
    \Call{RaiseToBatchedEinsum}{$K$}
    \State $e'_{\mathrm{ref}}, \sigma^{\mathcal{A}}_{\mathrm{canon-to-ref}},
    \sigma^{\mathcal{I}}_{\mathrm{canon-to-ref}} \gets$
    \Call{CanonicalizeBatchedEinsum}{$e_{\mathrm{ref}}$}
    \State $e'_{\mathrm{raised}}, \sigma^{\mathcal{A}}_{\mathrm{canon-to-raised}},
    \sigma^{\mathcal{I}}_{\mathrm{canon-to-raised}} \gets$
    \Call{CanonicalizeBatchedEinsum}{$e_{\mathrm{raised}}$}
    \State

    \If{$e'_{\mathrm{ref}} \neq e'_{\mathrm{raised}}$}
    \State Raise an exception.
    \EndIf
    \State

    \State $\sigma^{\mathcal{A}} \gets \{\}$
    \State $\sigma^{\mathcal{I}} \gets \{\}$
    \State

    \For{$a_{\mathrm{raised}}, a_{K} \in \sigma^{\mathcal{A}}_{\mathrm{raised}}$}
    \State
    $\sigma^{\mathcal{A}}[\sigma^{\mathcal{A}}_{\mathrm{canon-to-ref}}[(\sigma^{\mathcal{A}}_{\mathrm{canon-to-raised}})^{-1}[a_{\mathrm{raised}}]]]
    \gets a_K$
    \EndFor
    \For{$i_{\mathrm{raised}}, i_{K} \in \sigma^{\mathcal{I}}_{\mathrm{raised}}$}
    \State
    $\sigma^{\mathcal{I}}[\sigma^{\mathcal{I}}_{\mathrm{canon-to-ref}}[(\sigma^{\mathcal{I}}_{\mathrm{canon-to-raised}})^{-1}[i_{\mathrm{raised}}]]]
    \gets i_K$
    \EndFor

    \EndProcedure
  \end{algorithmic}
\end{algorithm}

\begin{minipage}{0.46\textwidth}
  \lstset{style=pythonstyle, frame=single, numbers=none,
  morekeywords={end}}
  \begin{lstlisting}[caption={Computational scientist's view of \textsc{Feinsum}.
    Querying the best performing transformation in \textsc{Feinsum}'s. database.},
                     label={lstng:demo_feinsum_subprogram}]
import loopy as lp
import feinsum as fs

knl = lp.make_kernel(
  "{ [i0,i1] : 0<=i0<96 and 0<=i1<4 }",
  """
  u(i, j) := P[i, j]*P[i, j]
  v(i) := 3*cos(Q[i])+5
  w(i) := sin(R[i])
  for i0
    y1[i0] = sum([i1], u(i0, i1)*v(i1))
    y2[i0] = sum([i1], u(i0, i1)*w(i1))
  end
  """)

transform = fs.retrieve(knl, device=...)
transformed_knl = transform(knl)\end{lstlisting}

\end{minipage}\hfill
\begin{minipage}{0.49\textwidth}

  \lstset{style=cstyle, frame=single, numbers=none,stepnumber=1}
\begin{lstlisting}[caption={\textsc{OpenCL} program for the subprogram
                            in Listing~\ref{lstng:demo_feinsum_subprogram}
                            transformed as per
                            Listing~\ref{lstng:demo_feinsum_transform}.},
                   label={lstng:demo_feinsum_opencl_subprogram}]
for (i0_out = 0; i0_out < 3; ++i0_out)
  for (i0_in = 0; i0_in < 32; ++i0_in) {
    double acc_i1 = acc_i1_0 = 0.0;
    for (int i1 = 0; i1 < 4; ++i1) {
      double p = P[i1+4*(32*i0_out+i0_in)]
               * P[i1+4*(32*i0_out+i0_in)];
      acc_i1   += p*(3*cos(Q[i1])+5);
      acc_i1_0 += p*(sin(R[i1]));
    }
    y1[32*i0_out + i0_in] = acc_i1;
    y2[32*i0_out + i0_in] = acc_i1_0;
  }
\end{lstlisting}
\end{minipage}

We now consider the computational scientist's perspective illustrated in
Listing~\ref{lstng:demo_feinsum_subprogram}. The computational scientist
expresses a functional batched einsum as a \textsc{Loopy} kernel and invokes the
\texttt{retrieve} routine to obtain the highest-performing transformation for
the corresponding idealized functional batched einsum stored in
\textsc{Feinsum}'s database.  During the call to \texttt{retrieve}, the kernel's
IR is traversed and raised to a functional batched einsum, which is subsequently
canonicalized and used to query the database.  The code generated as a result of
this retrieval is shown in Listing~\ref{lstng:demo_feinsum_opencl_subprogram},
assuming that the transformation in Listing~\ref{lstng:demo_feinsum_transform}
is the one retrieved. We highlight the transfer of the transformation, in
particular the loops that are tiled and the operands that are precomputed.

\section{Experimental Results}\label{sec:experiments}
In this section, we present a series of experiments designed to answer three
questions. First, are there real-world scenarios where using optimized
subprograms of batched einsums is profitable compared to utilizing a sequence of
optimized einsum subprograms? We investigate this in
Section~\ref{sec:perf_eval_dgfem} through a case study of batched einsum
expressions arising in a Discontinuous Galerkin Finite Element Method (DG-FEM)
solver. Second, is our choice of reusing optimizations for functional
batched einsums
relatively profitable over the traditional approach of storing executables
corresponding to optimized subprograms? We examine this by evaluating the
performance of the einsums from the TCCG suite~\cite{tccg2016a} in
Section~\ref{sec:perf_eval_tccg}. Finally, are the compile-time overheads
introduced in our \textit{identify-and-transform} strategy under control? We study
these code generation costs for the TCCG benchmark suite in
Section~\ref{sec:perf_eval_time_to_canonicalize}.

\subsection{Batched einsums in DG-FEM Operators}\label{sec:perf_eval_dgfem}
Discontinuous Galerkin Finite Element Method~\cite{hesthaven2007nodaldg}
(DG-FEM) is a popular numerical method for solving Partial Differential
Equations.  DG-FEM is attractive because it achieves high-order accuracy on
general grids while maintaining local conservation.  DG-FEM maps well to GPGPUs
as they have high cell-local computations and require relatively lower
cross-cell communication compared to other Finite Element Methods.

Kl\"{o}ckner et al.~\cite{klockner2009nodal} argue that three classes of batched
einsums are primarily responsible for most computations in a DG-FEM solver.
These batched einsums are:
\begin{description}
  \item[Face Mass Applications] These correspond to a class of kernels that
    can be raised to the batched einsums defined by the tuple $\bigl(b$, $3$,
      $(e,i)$, $\left(\left(f,e\right), \left(i,f,j\right),
      \left(f,e,j\right)\right)$, $\left(\left(M, J,
    F^k\right)\right)_{k=1}^n\bigr)$. Here, $F^k$ are the numerical flux fields,
    $M$ is the reference face mass matrix and $J$ contains the geometry terms.  $e$
    iterates over elements of the mesh, $f$ iterates over faces of a cell, $j$
    iterates over the degrees of freedom (DOFs) of the input fields that lie on
    the cell's faces, and, $i$ iterates over the output field's DOFs that lie
    inside the cell's volume.

    We show a kernel that implements a functional form of such batched einsums
    in Listing~\ref{lstng:feinsum_facemass_kernels}.
  \item[Local Divergence Computations] These correspond to a class of kernels that
    can be raised to the batched einsums that are
    defined by the tuple $\bigl(b$, $3$, $(e,i)$, $\left(\left(x,r,e\right),
      \left(r,i,j\right), \left(e,j\right)\right)$, $\left(\left(J, D,
    u^k\right)\right)_{k=1}^n\bigr)$. Here, $u^k$ are the input
    fields, $D$ is the reference derivative matrix and $J$ contains the geometry
    terms.  $e$ iterates over elements of the mesh, $x$ iterates over the
    topological dimension, $r$ iterates over the spatial dimension of the
    ambient function space, and $j, i$ iterate over the volume DOFs of the input
    and output fields respectively.

    We show a kernel that implements a functional form of such batched einsums
    in Listing~\ref{lstng:feinsum_divergence_kernels}.
  \item[Local Gradient Computations] These correspond to a class of kernels that
    satisfy can be identified as the batched einsums that are defined by the
    tuple $\bigl(b$, $3$, $(x, e,i)$, $\left(\left(x,r,e\right),
      \left(r,i,j\right), \left(x,e,j\right)\right)$, $\left(\left(J, D,
    u^k\right)\right)_{k=1}^n\bigr)$. Here, $u^k$ are the fields whose local
    gradients are to be computed, $D$ is the reference derivative matrix and $J$
    contains the geometry terms. $e$ iterates over elements of the mesh, $x$
    iterates over the topological dimension, $r$ iterates over the spatial
    dimension of the ambient function space, and $i, j$ iterate over the volume
    DOFs of the input and output fields respectively.

    We show a kernel that implements a functional form of such batched einsums
    in Listing~\ref{lstng:feinsum_gradient_kernels}.
\end{description}

We express the aforementioned batched einsums as \textsc{Loopy} kernels and
employ the transformations available in the default database integrated into our
\textsc{Feinsum} library. The transformation implementations for these batched
einsum operations employ a combination of loop permutation and loop hoisting to
reduce the total number of floating-point operations (FLOPs). In addition, we
apply a parametric tiling strategy for the matrices $J$ and $D$, together with a
parametric work-group grid. Parameter tuning is performed using
\textsc{OpenTuner}~\cite{ansel_2014_opentuner}. For brevity, we omit further
implementation details; interested readers are referred to Appendix~C of our
previous study~\cite{kulkarni2023domain} for a more comprehensive discussion.

As a baseline, we also implement the equivalent
operations using \textsc{JAX}, a state-of-the-art array library that leverages
lazy evaluation and supports operation fusion. The compilation heuristic in
\textsc{JAX} relies on \textsc{cuBLAS} and \textsc{cuDNN} for optimized
implementations for linear algebra primitives. In our \textsc{JAX}-based
implementation, we define a \textsc{Python} callable corresponding to the above
operations and decorate it with \texttt{jax.jit} to trace the array operations
within the function call.

All of our experiments were conducted on an Nvidia Titan V GPU, running the CUDA
driver version 550.163.  The transformed kernels are compiled using the Nvidia
\textsc{OpenCL} kernel compiler available in the CUDA toolkit version 12.4.

Given that our evaluation aims to assess the relative profitability of our
transformation strategies compared to state-of-the-art approaches, rather than
conducting an absolute performance evaluation, we report the speedups achieved
by the transformed kernels in comparison to the \textsc{JAX}
version. The speedup, denoted as $\eta$, is calculated as follows:
\[
  \eta = \frac{\text{Wall-clock time for the \textsc{JAX} kernels}}{\text{Wall-clock
  time for the transformed kernels}}.
\]

To avoid any transient effects, we compute the arithmetic mean of the
wall-times obtained by executing each kernel multiple times. Specifically, we
launch the kernel until the total wall-time spent reaches at least 2 seconds,
and the kernel is invoked at least 20 times. Prior to performing the
experiment for a kernel, we conduct 5 warm-up rounds to further reduce noise in
the experiments.

\begin{figure}
  \centering
  \begin{subfigure}[t]{0.45\textwidth}
    \centering
    \begin{tikzpicture}
      \begin{axis}[
          xlabel={Number of einsums $(b)$},
          ylabel={Speedup $(\eta)$},
          legend pos=outer north east,
          width=\textwidth
        ]
        \addplot[blue, mark=*, thick] table[x=x,y=ye1] {./face_mass_perf.dat};
        \addlegendentry{$e_1$}
        \addplot[red, mark=square*, thick] table[x=x,y=ye2]
        {./face_mass_perf.dat};
        \addlegendentry{$e_2$}
        \addplot[green!50!black,mark=triangle*, thick] table[x=x,y=ye3]
        {./face_mass_perf.dat};
        \addlegendentry{$e_3$}
        \addplot[orange, mark=+, mark options={line width=1.5pt}, thick]
        table[x=x,y=ye4] {./face_mass_perf.dat};
        \addlegendentry{$e_4$}
      \end{axis}
    \end{tikzpicture}
    \caption{Experimental results for Face Mass Applications. $e_k$
      corresponds to the batched einsum
      expressions in face-mass operators that are observed in DG-FEM solvers
      when employing a $P^k$ function space. For $e_1$, the index to axis
      length mapping is $\bigl\{i\mapsto4$; $e\mapsto10^5$; $f\mapsto4$;
      $j\mapsto3\bigr\}$.
      For $e_2$, the index to axis length mapping is
      $\bigl\{i\mapsto10$; $e\mapsto10^5$; $f\mapsto4$;
      $j\mapsto6\bigr\}$. For $e_3$,
      the index to axis length mapping is
      $\bigl\{i\mapsto20$; $e\mapsto10^5$; $f\mapsto4$;
      $j\mapsto10\bigr\}$. For $e_4$,
      the index to axis length mapping is
      $\bigl\{i\mapsto35$; $e\mapsto10^5$; $f\mapsto4$; $j\mapsto15\bigr\}$.
    }
  \end{subfigure}\hspace{.09\textwidth}%
  \begin{subfigure}[t]{0.45\textwidth}
    \centering
    \begin{tikzpicture}
      \begin{axis}[
          xlabel={Number of einsums $(b)$},
          ylabel={Speedup $(\eta)$},
          legend pos=outer north east,
          width=\textwidth
        ]
        \addplot[blue, mark=*, thick] table[x=x,y=ye1] {./divergence_perf.dat};
        \addlegendentry{$e_1$}
        \addplot[red, mark=square*, thick] table[x=x,y=ye2]
        {./divergence_perf.dat};
        \addlegendentry{$e_2$}
        \addplot[green!50!black,mark=triangle*, thick] table[x=x,y=ye3]
        {./divergence_perf.dat};
        \addlegendentry{$e_3$}
        \addplot[orange, mark=+, mark options={line width=1.5pt}, thick]
        table[x=x,y=ye4] {./divergence_perf.dat};
        \addlegendentry{$e_4$}
      \end{axis}
    \end{tikzpicture}
    \caption{Experimental results for Local Divergence Computations.
      $e_k$ corresponds to the batched einsum
      expressions in local-divergence operators that are observed in DG-FEM
      solvers when employing a $P^k$ function space. For $e_1$, the index to
      axis length mapping is
      $\bigl\{i\mapsto4$; $j\mapsto4$; $e\mapsto10^5$; $x\mapsto3$;
      $r\mapsto3\bigr\}$.
      For $e_2$, the index to axis length mapping is
      $\bigl\{i\mapsto10$; $j\mapsto10$; $e\mapsto10^5$; $x\mapsto3$;
      $r\mapsto3\bigr\}$.
      For $e_3$, the index to axis length mapping is
      $\bigl\{i\mapsto20$; $j\mapsto20$; $e\mapsto10^5$; $x\mapsto3$;
      $r\mapsto3\bigr\}$.
      For $e_4$, the index to axis length mapping is
      $\bigl\{i\mapsto35$; $j\mapsto35$; $e\mapsto10^5$; $x\mapsto3$;
    $r\mapsto3\bigr\}$.}
  \end{subfigure}
  \begin{subfigure}[t]{.45\textwidth}
    \centering
    \begin{tikzpicture}
      \begin{axis}[
          xlabel={Number of einsums $(b)$},
          ylabel={Speedup $(\eta)$},
          legend pos=outer north east,
          width=\textwidth
        ]
        \addplot[blue, mark=*, thick] table[x=x,y=ye1] {./grad_perf.dat};
        \addlegendentry{$e_1$}
        \addplot[red, mark=square*, thick] table[x=x,y=ye2] {./grad_perf.dat};
        \addlegendentry{$e_2$}
        \addplot[green!50!black,mark=triangle*, thick] table[x=x,y=ye3]
        {./grad_perf.dat};
        \addlegendentry{$e_3$}
        \addplot[orange, mark=+, mark options={line width=1.5pt}, thick]
        table[x=x,y=ye4] {./grad_perf.dat};
        \addlegendentry{$e_4$}
      \end{axis}
    \end{tikzpicture}
    \caption{ Experimental results for Local Gradient Computations. $e_k$
      corresponds to the batched einsum expressions in local-divergence
      operators that are observed in DG-FEM solvers when employing a $P^k$
      function space. For $e_1$, the index to axis length mapping is
      $\bigl\{i\mapsto4$; $j\mapsto4$; $e\mapsto10^5$; $x\mapsto3$;
      $r\mapsto3\bigr\}$. For $e_2$, the index to axis length mapping is
      $\bigl\{i\mapsto10$; $j\mapsto10$; $e\mapsto10^5$; $x\mapsto3$;
      $r\mapsto3\bigr\}$. For $e_3$, the index to axis length mapping is
      $\bigl\{i\mapsto20$; $j\mapsto20$; $e\mapsto10^5$; $x\mapsto3$;
      $r\mapsto3\bigr\}$. For $e_4$, the index to axis length mapping is
      $\bigl\{i\mapsto35$; $j\mapsto35$; $e\mapsto10^5$; $x\mapsto3$;
    $r\mapsto3\bigr\}$.}
  \end{subfigure}
  \caption{Observed Speedups for batched einsums executed using
    \textsc{Feinsum} relative to equivalent implementation using
  \textsc{JAX}/.}\label{fig:feinsum_speedup}
\end{figure}
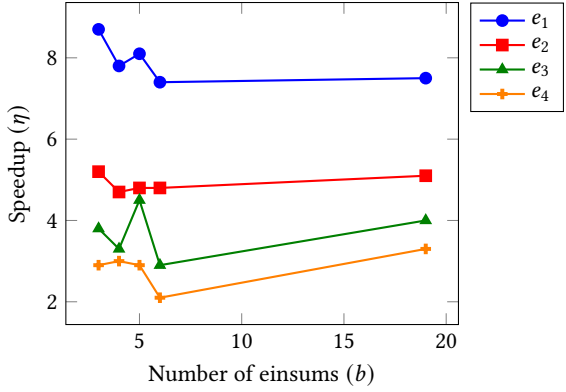
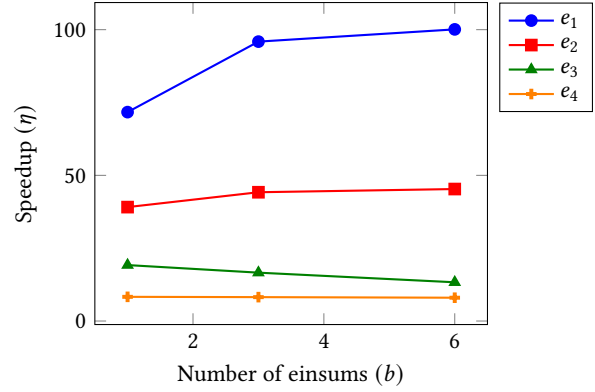
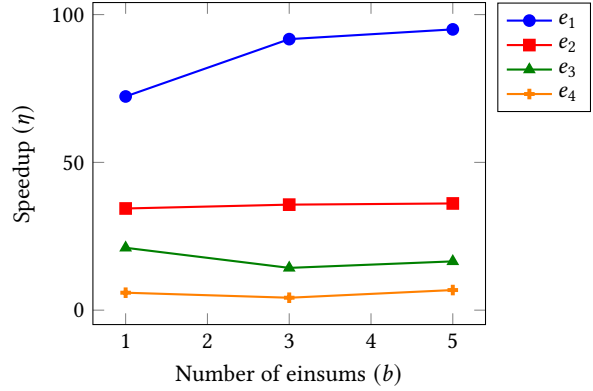

We present the obtained speedup values in Figure~\ref{fig:feinsum_speedup}. The
charts indicate a speedup in the range of $1.6$ to $100$ is obtained for our
suite of macro-kernels, affirming the benefits offered by our transformation
abstraction over the state-of-the-art approaches. We expect higher speedup
values as the number of einsums in a batched einsum increases, as relatively
this leads to higher cost savings in loading the data on the GPU's DRAM.
However, it should be noted that this trend is not consistently observed in the
data. This is primarily because our database does not contain all the
transformations capable of achieving roofline-level performance for every
einsum operation, thereby deviating from the ideal trend.

This experiment supports our claim that using batched einsums as a primitive
allows the computational implementation to deliver higher performance by hiding
latency corresponding to the memory accesses of operands shared across the
einsums.

\subsection{TCCG Benchmark suite with Unmaterialized
Operands}\label{sec:perf_eval_tccg}
Springer et al.~\cite{tccg2016a} compiled a series of 48 tensor contractions
seen across a broad range of use cases such as electronic structure
computations, coupled cluster theory, and quantum chemistry. To model a real
world scenario, we propose a slight extension to the benchmark suite, where
instead of the operands to the tensor contractions being materialized operands
we consider the operands be linear polynomials of arrays in memory.
Specifically, instead of evaluating the performance of tensor contractions with
the arrays $A$ and $B$, we evaluate the performance of the tensor contraction
with the operands, $\left(\alpha_1 A + \beta_1\right)$ and $\left(\alpha_2 B +
\beta_2\right)$, where $\alpha_1$, $\alpha_2$, $\beta_1$, $\beta_2$ are scalars
only known at runtime. In all our benchmarks, $\alpha_1$, $\alpha_2$, $\beta_1$,
and $\beta_2$ are double-precision scalars and $A$ and $B$ are multidimensional
arrays that have double-precision entries.

For each test case in the test suite, we report the FLOP throughput as
\[
  \frac{\text{FLOPS in the contraction from \textsc{Opt-einsum}} +
    \text{FLOPS for computing }\alpha_1 A + \beta_1 \text{ and } \alpha_2 B +
  \beta_2}{t_{\text{wall}}},
\]
where $t_{\text{wall}}$ is the averaged wall-clock time spent in the kernel. To
obtain $t_{\text{wall}}$, we perform 5 warmup rounds and use the arithmetic mean
of wall-time for the execution of next $N$ rounds, where $N$ is the smallest
number greater than 20 such that the aggregate wall-time spent inside the
kernel is at least 2 seconds. We perform these tests on an Nvidia Titan V GPU,
running the CUDA driver version 550.163. The transformed kernels are compiled
using the Nvidia \textsc{OpenCL} kernel compiler available in the CUDA
toolkit version
12.4.

We also report the roofline performance~\cite{williams_2009_roofline} via a
conservative model assuming that the GPU is a latency hiding machine whose
performance can be limited either by the execution units or memory units. Hence,
we arrive at:
\[
  \text{Roofline FLOPS} = \min(\text{Peak FLOPS}, \frac{\text{Total
  FLOPS}}{\text{\# Footprint Bytes}} \cdot \text{Peak Bandwidth}).
\]
We use the manufacturer-reported values for peak FLOPS and bandwidth values.

\pgfplotstableread[col sep=space]{./tccg_perf_feinsum_jax.dat}\tccgperfdatatable
\begin{figure}[h]
  \centering
  \begin{tikzpicture}
    \begin{axis}[
        xtick=data,
        xticklabels from table={\tccgperfdatatable}{xticklabel},
        xticklabel style={rotate=90},
        ylabel={GFLOPS},
        xticklabel style = {font=\tiny},
        ylabel style = {font=\small},
        xmin=0.5,
        xmax=48.5,
        width=1\linewidth,
        height=0.4\linewidth,
        legend pos= north east,
        legend style= {font=\small}
      ]
      \addplot+[
        thick,
        color=black,
        dashed,
        mark=none
      ] table[x=x, y=yroofline]{\tccgperfdatatable};
      \addlegendentry{Roofline}

      \addplot+[
        thick,
        color=blue!70!black,
        mark=*,
        mark options={scale=1.1}
      ] table[x=x, y=yfeinsum]{\tccgperfdatatable};
      \addlegendentry{Feinsum}

      \addplot+[
        thick,
        color=red!70!black,
        mark=square,
        mark options={scale=1.1}
      ] table[x=x, y=yjax]{\tccgperfdatatable};
      \addlegendentry{JAX}
    \end{axis}
  \end{tikzpicture}
  \caption{FP64 FLOP throughput for the benchmarks based on the
  TCCG-suite as detailed in Section~\ref{sec:perf_eval_tccg}.}
  \label{fig:perf_eval_tccg}
\end{figure}
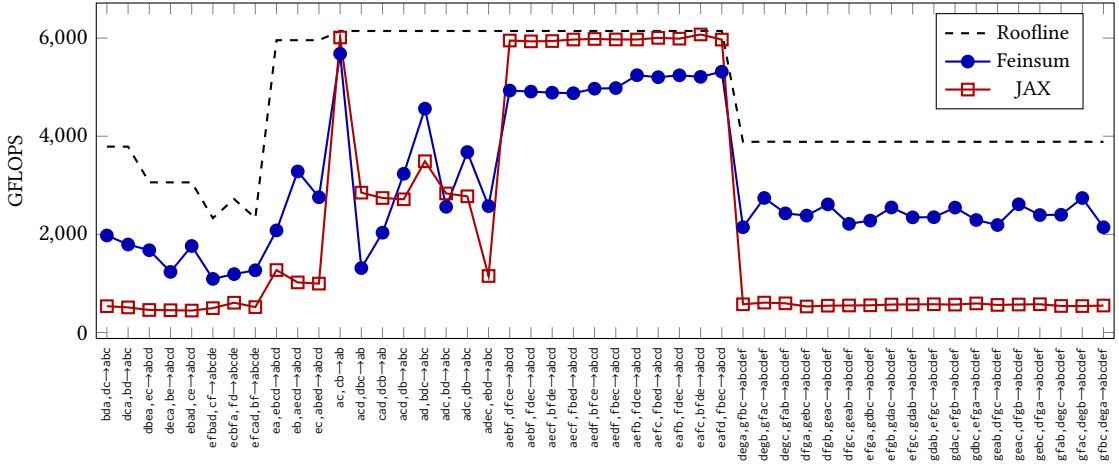

We present the data from these experiments in Figure~\ref{fig:perf_eval_tccg}.
It can be observed that our transformation database contains entries that
deliver $2.1\times$ geometric-mean speedup over JAX for the TCCG suite. We
attribute this speedup to two key reasons. First, JAX uses a transformation
strategy similar to TTGT~\cite{hirata2003tensor} and relies on the
matrix-multiply operation available in \textsc{cuBLAS}, and our transformation
database contains entries based on TTGT and the COGENT~\cite{kim2019code}
transformation space. COGENT was previously shown to be profitable over TTGT in
a geometric mean sense. Second, our \emph{identify-and-transform} strategy
couples that with fusing
surrounding computations for the operands. For memory bound test-cases,
cost of the actual floating point should not be apparent in wall-clock timings.
This is evident by the fact that the relative speedup of our work over JAX is
$3.6\times$ for the memory-bound cases of Figure~\ref{fig:tccg_ai}.

This experiment validates our claim that the reliance of widely used array
compilers on \textsc{BLAS} primitives significantly limits their performance of
tensor contraction expressions in practical programs. Meanwhile, our effort to
tabulate optimized transformations for idealized batched einsums shows
substantial cost reductions across this benchmark suite.

\subsection{Time to canonicalize a batched
einsum}\label{sec:perf_eval_time_to_canonicalize}

We also aim to analyze the cost associated with the transformations, which is
not accounted for in the kernel execution time.  While these overheads are
observed during the compilation of a larger program whose executable is reused
multiple times, high compile times overheads can potentially impact the
practicality of the proposed system. In our approach, these overheads include
the batched einsum canonicalization process and the AST traversal for matching
an einsum. Among these operations, matching an expression involves inexpensive
AST traversals. However, batched einsum canonicalization relies on graph
canonicalization, which exhibits sub-exponential complexity in terms of the
number of nodes in the graph. This motivates us to conduct another series of
studies to quantify the expenses incurred during the canonicalization process.

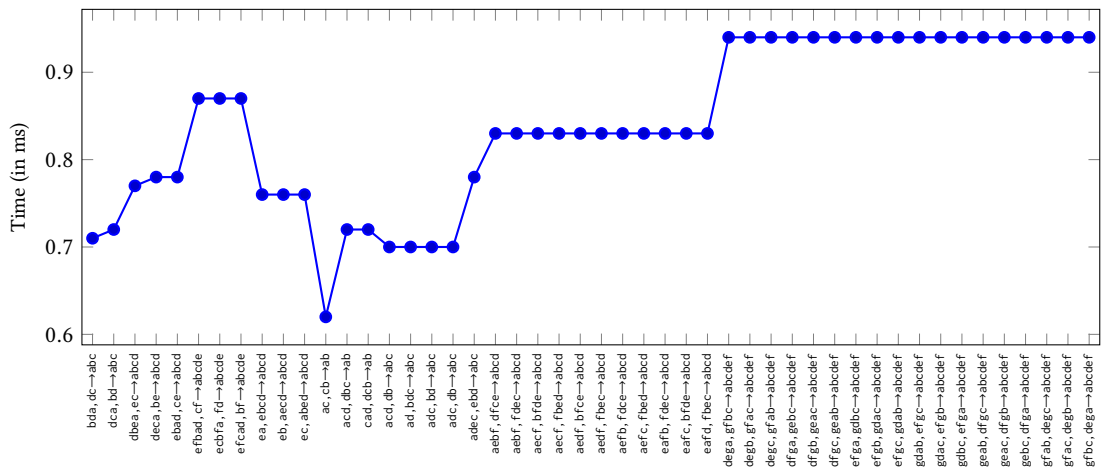
\begin{figure}[h]
  \centering
  \begin{tikzpicture}
    \begin{axis}[
        xtick=data,
        xticklabels from table={\tccgcanonoverheaddatatable}{xticklabel},
        xticklabel style={rotate=90},
        ylabel={Time (in ms)},
        xticklabel style = {font=\tiny},
        ylabel style = {font=\small},
        xmin=0.5,
        xmax=48.5,
        width=1\linewidth,
        height=0.4\linewidth
      ]
      \addplot+[thick] table[x=x, y=y]{\tccgcanonoverheaddatatable};
    \end{axis}
  \end{tikzpicture}
  \caption{Canonicalization costs for the TCCG benchmark
  suite.}\label{fig:feinsum_canon_time_plot}
\end{figure}

In Figure~\ref{fig:feinsum_canon_time_plot}, we present the wall-clock times
required to canonicalize an instance of each einsum in the TCCG benchmark. These
experiments were conducted on an AMD Ryzen 7 5800HS CPU. Across the benchmark
suite, we consistently observe that the canonicalization process takes less
than 1 millisecond.

\pgfplotstableread[col
sep=space]{./tccg_canonicalization_overheads.dat}\tccgcanonoverheaddatatable

\subsection{Code Availability}
To ensure reproducibility, we provide a DOI for an archived version of the
software used to generate the results presented in this article. The software
toolkit containing the canonicalization algorithm from
Section~\ref{sec:canonicalization_via_directed} and an implementation of the
abstraction of Section~\ref{sec:feinsum_abstraction} have been provided as
\textsc{Python} library which have been archived on
Zenodo~\cite{zenodo/feinsum_repo_2025} at the versions studied here. Similarly,
the software utilized to obtain the empirical data in
Section~\ref{sec:experiments} has been archived on
Zenodo~\cite{zenodo/feinsum_eval_2025}. Detailed installation instructions are
provided in the README file included in the archived records.  All contents of
the archive are distributed under the MIT License.

\section{Conclusions}
In this work, we have presented three contributions. First, we have proposed an
extension to the einsum computational primitive to support batches of them.
Batching is an important and a commonly overlooked aspect in the synthesis of
efficient code for dense linear-algebra programs. We have provided a series of
experiments in the context of real-world applications to support this claim.
Second, we generalize the concept of batched einsums to a formulation that
accepts functions as operands, thereby facilitating operator fusion for
optimizations known from their idealized versions.  The case study on a modified
TCCG benchmark suite has highlighted the need for such techniques in efficient
handling of memory bound batched einsums.  Finally, we have implemented an
open-source software system that leverages the aforementioned abstractions to
maintain a database of optimized code-transformations for the class of batched
einsum expressions.

This works leads to some interesting questions for future research:

\begin{description}

  \item[Is there an einsum canonicalization algorithm with polynomial complexity?]
    In this work, we use graph canonicalization as an intermediate step for
    the canonicalization of einsums. However, an algorithm for graph
    canonicalization with a worst-case bound on asymptotic complexity that
    is polynomial is currently unavailable.  One possibility is to employ Babai et
    al.'s~\cite{babai2019canonical} graph canonicalization algorithm to achieve
    an algorithm with quasi-polynomial complexity. However, it remains an open
    question whether an einsum canonicalization algorithm exists that does not
    rely on graph canonicalization as an intermediate step.

  \item[What is the tight upper bound for the performance of an einsum
    expression?]
    A roofline model specific to the class of batched einsums is currently
    unavailable. The performance of einsum-like expressions depends on the
    device's memory hierarchy, as the same memory location is accessed multiple
    times. This makes it challenging to develop a model that accounts for all
    possible bottlenecks, including L1 cache and local memory. While obtaining a
    model based on global memory bandwidth is relatively straightforward, it
    does not capture all the performance considerations in this context.  Such a
    model could be helpful to transform writers to inform them on when to expect
    a tuning process to yield diminishing returns.
\end{description}

\begin{acks}
  The authors' research was funded by the US National Science Foundation under
  awards SHF-1911019 and OAC-1931577, by the US Department of Energy under award
  number DE-NA0003963, as well as by the Siebel School of Computing and Data
  Science at the University of Illinois at Urbana-Champaign. Any opinions,
  findings, and conclusions, or recommendations  expressed in this article are
  those of the authors and do not necessarily reflect the views of the sponsors;
  the sponsors have not approved or endorsed its content.
\end{acks}

\bibliographystyle{ACM-Reference-Format}
\bibliography{references}

\appendix

\section{An overview of \textsc{Loopy} Intermediate
Representation}\label{sec:preliminaries_loopy}
\textsc{Loopy} Intermediate Representation (IR), introduced by Kl\"{o}ckner et
al.~\cite{klockner2014loo}, is a mid-level IR for modeling and
transforming array-based programs. In the \textsc{Loopy} IR, a key data
structure is the \textit{kernel}, which consists of multiple statements
executed within a polyhedrally defined iteration domain. These statements
operate on arrays with specific memory layouts, data types, and address spaces.
Additionally, a \textsc{Loopy} kernel encodes a dependence graph among the
statements, which provides a partial specification of the dependencies. We
present an example of a \textsc{Loopy} kernel in
Listing~\ref{lstng:feinsum_loopy_demo}.

\begin{center}
  \begin{minipage}{47ex}
    \centering
    \lstset{style=pythonstyle, frame=single, numbers=left,stepnumber=1,
    morekeywords={INSTRUCTIONS,DOMAINS,ARGUMENTS, SUBSTITUTION, RULES,end}}
  \begin{lstlisting}[
                     caption={A \textsc{Loopy} kernel evaluating arrays
                              \texttt{y1} and \texttt{y2}.},
                     label={lstng:feinsum_loopy_demo}]
 ARGUMENTS:
 A: dtype: float64, shape: (n, n)
 n: ValueArg, dtype: int32
 x: dtype: float64, shape: (n)
 y1: dtype: float64, shape: (n)
 y2: dtype: float64, shape: (n)
 --------------------------------
 DOMAINS:
 [n] -> { [i, j] : 0 <= i,j < n }
 --------------------------------
 SUBSTITUTION RULES:
 u1(i) := 2*cos(x[i]) + 3
 u2(i) := 3*sin(x[i]) + 5
 --------------------------------
 INSTRUCTIONS:
 for i
   y1[i] = sum([j], A[i, j]*u1(j))
   y2[i] = sum([j], A[i, j]*u2(j))
 end i\end{lstlisting}
  \end{minipage}
\end{center}

Additionally, \textsc{Loopy} IR provides a node called a \textit{substitution
rule}. These are declared on lines 12--13 of the listing and invoked on lines
17--18. These substitution rules~\cite{klockner2015substitutions} are inlined
during code-generation, however, they can be used to implement targeted
transformations such as, prefetching the accesses to the substitution rule to a
local / private temporary. Kl\"{o}ckner et al.~\cite{klockner2016loopyfem}
demonstrated this in the context of higher order Finite Element Methods.

In Section~\ref{sec:feinsum_abstraction}, we present a specification for storing
transformation knowledge for programs that execute einsum operations.  In this
specification, we use \textsc{Loopy} both as a descriptive grammar and as a
language for specifying transformations. Notably, the contributions in this paper
are independent of \textsc{Loopy}, which is used solely as an IR and a
transformation description language for the accompanying implementation.

\section{Induced Graph}
In this section, we provide a mapping from a batched einsum to a colored graph
and a mapping from a colored graph to a batched einsum. In
Section~\ref{sec:canonicalization_via_directed}, we saw that these mappings are
fundamental to our batched canonicalization algorithm. Prior to describing our
graph constructions corresponding to batched einsums, we provide the definitions
of the following terms which will be helpful during the process.

\begin{definition}[All dimensions in an einsum]
  We define the set of all dimensions for a batched einsum $e$ as
  \[
    \operatorname{AllDims}(e) = \{\operatorname{Dim}(a):\;\; a \in
    \mathcal{A}^{\mathrm{all}}(e)\} \cup \{|\mathcal{I}^{\mathrm{out}}(e)|\}.
  \]
\end{definition}

\begin{definition}[Input accesses in an einsum]
  We define the set of all the input accesses for a batched einsum $e$ as
  \begin{equation}
    \begin{array}{ll}
      \operatorname{InputAccesses}(e) = \bigcup\limits_{j=1}^{n}\Bigl\{ & (i,
        j, \mathcal{I}^{\mathrm{in},j}_d, d):\\
        & d\in\left\{1, \ldots, |{\mathcal{I}}^{\mathrm{in},j}|\right\},
      i\in\{1,2,\ldots,b\}, j\in\{1,2,\ldots,n\}\Bigr\}
    \end{array}
  \end{equation}
\end{definition}

\begin{definition}[Output accesses in an einsum]
  We define the set of all the output accesses for a batched einsum $e$ as
  \[
    \operatorname{OutputAccesses}(e) =
    \left\{\left(\mathcal{I}^{\mathrm{out}}_d, d\right):
    d\in\{1,\ldots,|\mathcal{I}^{\mathrm{out}}|\}\right\}.
  \]
\end{definition}

\begin{definition}[All data types in a batched einsum]
  The set of data types of the inputs in a batched einsum, $e$, is given by
  \[
    {\operatorname{Dtypes}}(e) = \{{\operatorname{Dtype}}(a): a\in
    \mathcal{A}^{\mathrm{all}}(e)\}.
  \]
\end{definition}

\begin{definition}[Axis lengths in a batched einsum]
  The set of axis lengths in a batched einsum, $e$, is given by
  \[
    {\operatorname{AxisLengths}}(e) = \left\{
      \left({\operatorname{Shape}}(a)\right)_i: i\in \left\{1, \ldots,
      \operatorname{Dim}(a)\right\}, a\in \mathcal{A}^{\mathrm{all}}(e)
    \right\}.
  \]
\end{definition}
\begin{definition}[Induced Graph]\label{defn:induced_directed_graph}
  A graph induced by a batched einsum
  $\bigl(N^{\mathrm{IDG}}$, $\iota_{\mathrm{dtype}}$, $\iota_{\mathrm{lengths}}$,
  $A^{N^{\mathrm{IDG}} \times N^{\mathrm{IDG}}}$, $c_{N^{\mathrm{IDG}}}\bigr)$,
  where:
  \begin{itemize}
    \item $N^{\mathrm{IDG}}$ is the number of nodes in the graph
    \item $A$ is the adjacency matrix of the graph.
    \item $c$ is the coloring vector of the nodes in the graph.
    \item $\iota_{\mathrm{dtype}}$ is a mapping from a subset
      of $\{1,2,\ldots,N^{\mathrm{IDG}}\}$ to numeric data types.
    \item $\iota_{\mathrm{length}}$ is a mapping from a subset of
      $\{1,2,\ldots,N^{\mathrm{IDG}}\}$ to non-negative integers.
  \end{itemize}
\end{definition}

\subsection{Mapping Batched Einsum to Induced
Graph}\label{sec:to_directed_induced_graph}
We proceed to outline our construction process for an \textit{induced
graph} for a batched einsum $e$ defined by the tuple $\Bigl(b$, $n$,
  $\mathcal{I}^{\mathrm{out}}$, $(\mathcal{I}^{\mathrm{in},k})_{k=1}^{n}$,
$\left(\left(\mathcal{A}^{j, k}\right)_{k=1}^{n}\right)_{j=1}^b\Bigr)$.
We will use the following notation to simplify our construction process:
\begin{itemize}
  \item $N_{\mathrm{arg}} = |\mathcal{A}^{\mathrm{all}}(e)|$
  \item $N_{\mathrm{index}} = |\mathcal{I}^{\mathrm{all}}(e)|$
  \item $N_{\mathrm{length}} = |\operatorname{AxisLengths}(e)|$
  \item $N_{\mathrm{dim}} = \max\left(\operatorname{Dims}(e)\right)$
  \item $N_{\mathrm{dtypes}} = |\operatorname{Dtypes}(e)|$
  \item $N_{\mathrm{access,in}} = |\operatorname{InputAccesses}(e)|$
  \item $N_{\mathrm{access,out}} = |\operatorname{OutputAccesses}(e)|$
\end{itemize}

In this construction, we assign nodes to various components of a
batched einsum as follows: one node per argument, one node per index, one
node per unique numeric data type of the arrays, one node per unique axis
length, one node per index access, one node per dimension index, and one node
per output. Consequently, we obtain the number of nodes in the graph as
follows:
\begin{equation}\label{eq:directed_induced_graph_n_idg}
  N^{\mathrm{IDG}} =
  \Bigl(N_{\mathrm{arg}} + N_{\mathrm{index}}
    + N_{\mathrm{dtype}} + N_{\mathrm{length}}\\
    + N_{\mathrm{dim}}  + N_{\mathrm{access, in}} + N_{\mathrm{access,
  out}}  + b + n\Bigr)
\end{equation}

We now define the mappings from the different components of the batched
einsum to the node labels in our graph.

\begin{itemize}[label={$\text{-}$}]
  \item Choose a numbering for the arrays, denoted by the bijective map
    $\iota_{\mathrm{arg}}: \{1, 2, \ldots,
    N_{\mathrm{arg}}\} \mapsto \mathcal{A}^{\mathrm{all}}$.

  \item Choose a numbering for the indices, denoted by the bijective map
    $\iota_{\mathrm{index}}:
    \{N_{\mathrm{arg}}+1,
    \ldots, N_{\mathrm{arg}}+N_{\mathrm{index}}\} \mapsto
    \mathcal{I}^{\mathrm{all}}$.

  \item Choose a numbering for the input accesses, denoted by the bijective map
    $\iota_{\mathrm{access,in}}:
    \{N_{\mathrm{arg}}+N_{\mathrm{index}}+1,
    \ldots, N_{\mathrm{arg}}+N_{\mathrm{index}}+N_{\mathrm{access,in}}\}
    \mapsto \operatorname{InputAccesses}(e)$.

  \item Choose a numbering for the output accesses, denoted by the bijective
    map
    $\iota_{\mathrm{access,out}}:
    \{N_{\mathrm{arg}}+N_{\mathrm{index}}+N_{\mathrm{access,in}}+1,
      \ldots,
    N_{\mathrm{arg}}+N_{\mathrm{index}}+N_{\mathrm{access,in}}+N_{\mathrm{access,out}}\}
    \mapsto \operatorname{OutputAccesses}(e)$.

  \item Choose a numbering for the outputs, denoted by the bijective map
    $\iota_{\mathrm{output}}:
    \{N_{\mathrm{arg}}+N_{\mathrm{index}}+N_{\mathrm{access,in}}+N_{\mathrm{access,out}}+1,
      \ldots,
    N_{\mathrm{arg}}+N_{\mathrm{index}}+N_{\mathrm{access,in}}+N_{\mathrm{access,out}}+b\}
    \mapsto \{1, 2, \ldots, b\}$.

  \item Choose a numbering for the operand positions, denoted by the bijective
    map
    $\iota_{\mathrm{arg-pos}}:
    \{N_{\mathrm{arg}}+N_{\mathrm{index}}+N_{\mathrm{access,in}}+N_{\mathrm{access,out}}+b+1,
      \ldots,
    N_{\mathrm{arg}}+N_{\mathrm{index}}+N_{\mathrm{access,in}}+N_{\mathrm{access,out}}+b+n\}
    \mapsto \{1, 2, \ldots, n\}$.

  \item Choose a numbering for the data types,
    denoted by the bijective map
    \begin{equation}\label{eq:directed_induced_graph_iota_dtype}
      \begin{array}{ll}
        \iota_{\mathrm{dtype}}:
        \bigl\{&N_{\mathrm{arg}}+N_{\mathrm{index}}+N_{\mathrm{access,in}}+N_{\mathrm{access,out}}+b+n+1,\\
          & \ldots, \\
          &
        N_{\mathrm{arg}}+N_{\mathrm{index}}+N_{\mathrm{access,in}}+N_{\mathrm{access,out}}+b+n+N_{\mathrm{dtype}}\bigr\}
        \mapsto \operatorname{Dtypes}(e).
      \end{array}
    \end{equation}

  \item Choose a numbering for the axis lengths,
    denoted by the bijective map
    \begin{equation}\label{eq:directed_induced_graph_iota_length}
      \begin{array}{l}
        \iota_{\mathrm{length}}:
        \bigl\{ \\
          \quad\qquad
          N_{\mathrm{arg}}+N_{\mathrm{index}}+N_{\mathrm{access,in}}+N_{\mathrm{access,out}}+b+n+N_{\mathrm{dtype}}+1,\\
          \quad\qquad \ldots, \\
          \quad\qquad
        N_{\mathrm{arg}}+N_{\mathrm{index}}+N_{\mathrm{access,in}}+N_{\mathrm{access,out}}+b+n+N_{\mathrm{dtype}}+N_{\mathrm{length}}\bigr\}\mapsto\operatorname{AxisLengths}(e).
      \end{array}
    \end{equation}

  \item Choose a numbering for the dimensions of an array, denoted by the
    bijective map,
    \begin{equation}
      \begin{array}{rl}
        \iota_{\mathrm{dim}}=\{&\left(N_{\mathrm{arg}}+N_{\mathrm{index}}+N_{\mathrm{access,in}}+N_{\mathrm{access,out}}+b+n+N_{\mathrm{dtype}}+N_{\mathrm{length}}+i\right)\mapsto
          i:\\
        &i\in \{1, 2, \ldots, N_{\mathrm{dim}}\}\}
      \end{array}
    \end{equation}
\end{itemize}

We proceed to define specific sets that contribute to the edges in the graph,
representing the interactions among various components in $e$.

\begin{itemize}[label=\textendash]
  \item We define a set of edges to characterize the indexing
    contribution of an input
    array to a output. We denote this as $E_{\mathrm{access,in-to-arg}}$.
    \begin{eqnarray*}
      E_{\mathrm{access,in-to-arg}} = \Bigl\{
        &\left(\iota^{-1}_{\mathrm{access,in}}\left(\left(\left(i, j,
              \mathcal{I}^{\mathrm{in},j}\right)_d,
        d\right)\right),\iota^{-1}_{\mathrm{arg}}\left(\mathcal{A}^{i,j}\right)\right):\\
        & d\in \{1,\ldots,
        \operatorname{Dim}(\mathcal{A}^{i,j})\},
        j \in \{1, \ldots, n\},
      i \in \{1, \ldots, b\}\Bigr\}
    \end{eqnarray*}

  \item We now define edges that characterize an access node i.e. its index,
    the dimension being indexed over and the output it is contributing to. We
    denote these as $E_{\mathrm{arg-pos-to-access,in}}$,
    $E_{\mathrm{output-to-access,in}}$, $E_{\mathrm{index-to-access,in}}$,
    and, $E_{\mathrm{dim-to-access,in}}$.
    \begin{align}
      E_{\mathrm{arg-pos-to-access,in}} = \{(\iota^{-1}_{\mathrm{arg-pos}}(a),
        \iota^{-1}_{\mathrm{access,in}}\left((j, a, i, k)\right)):
      (j, a, i, k)\in \mathrm{InputAccesses}(e)\} \\
      E_{\mathrm{output-to-access,in}} = \{(\iota^{-1}_{\mathrm{output}}(i),
        \iota^{-1}_{\mathrm{access,in}}\left((j, a, i, k)\right)):
      (j, a, i, k)\in \mathrm{InputAccesses}(e)\} \\
      E_{\mathrm{index-to-access,in}} = \{(\iota^{-1}_{\mathrm{index}}(j),
        \iota^{-1}_{\mathrm{access,in}}\left((j, a, i, k)\right)):
      (j, a, i, k)\in \mathrm{InputAccesses}(e)\} \\
      E_{\mathrm{dim-to-access,in}} = \{(\iota^{-1}_{\mathrm{dim}}(k),
        \iota^{-1}_{\mathrm{access,in}}\left((j, a, i, k)\right)):
      (j, a, i, k)\in \mathrm{InputAccesses}(e)\}
    \end{align}

  \item We now define edges that characterize an output indexing node. We
    denote these as $E_{\mathrm{index-to-access,out}}$, and,
    $E_{\mathrm{dim-to-access,out}}$.
    \begin{align}
      E_{\mathrm{index-to-access,out}} = \{(\iota^{-1}_{\mathrm{index}}\left(i\right),
        \iota^{-1}_{\mathrm{access,out}}\left((i, d)\right)):
      (i, d)\in \mathrm{OutputAccesses}(e)\} \\
      E_{\mathrm{dim-to-access,out}} = \{(\iota^{-1}_{\mathrm{dim}}\left(d\right),
        \iota^{-1}_{\mathrm{access,out}}\left((i, d)\right)):
      (i, d)\in \mathrm{OutputAccesses}(e)\}
    \end{align}

  \item We introduce edges to characterize the iteration count corresponding to
    each index. We denote this as $E_{\mathrm{length}}$:
    \begin{equation}
      \begin{array}{ll}
        E_{\mathrm{length}} = \bigl\{ &
          \left(\iota^{-1}_{\mathrm{length}}\left(\left(\operatorname{Shape}(\mathcal{A}^{i,k})\right)_d\right),
          \iota^{-1}_{\mathrm{index}}\left((\mathcal{I}^{\mathrm{in},k})_d\right)\right):\\
          & d \in \{1,\ldots,\operatorname{Dim}(\mathcal{A}^{i,k})\}, k \in
        \{1,\ldots,n\}, i\in\{1, \ldots, b\} \bigr\}\\
      \end{array}
    \end{equation}

  \item We introduce edges to characterize the numeric data type corresponding
    to each array. We denote this as $E_{\mathrm{dtype}}$.
    \[
      E_{\mathrm{dtype}} = \left\{
        \left(\iota^{-1}_{\mathrm{dtype}}\left(\operatorname{Dtype}(a)\right),
        \iota^{-1}_{\mathrm{arg}}(a)\right):
        \;\; a \in \mathcal{A}^{\mathrm{all}}(e)
      \right\}
    \]

  \item We introduce edges to record the difference between the axis length
    nodes. We denote this as $E_{\mathrm{length ranks}}$.
    \begin{equation}
      \label{eq:to_induced_graph_length_ranks}
      E_{\mathrm{length ranks}} =
      \left\{(\iota^{-1}_{\mathrm{length}}\left(l_i\right),
        \iota^{-1}_{\mathrm{length}}\left(l_j\right)): l_j > l_i,
        l_i\in\operatorname{AxisLengths}(e),
      l_j\in\operatorname{AxisLengths}(e)\right\}
    \end{equation}

  \item We introduce edges to record the difference between the data type
    nodes. We denote this as $E_{\mathrm{dtype ranks}}$.
    \begin{equation}
      \label{eq:to_induced_graph_dtype_ranks}
      \begin{array}{ll}
        E_{\mathrm{dtype ranks}} = \bigl\{ &
          (\iota^{-1}_{\mathrm{dtype}}\left(t_i\right), \iota^{-1}_{\mathrm{dtype}}\left(t_j\right)): \\
          & \operatorname{DtypeRank}\left(t_j\right) > \operatorname{DtypeRank}\left(t_i\right),\\
          & t_i\in\operatorname{Dtypes}(e),\\
        & t_j\in\operatorname{Dtypes}(e)\bigr\}\\
      \end{array}
    \end{equation}
    where, \textsc{DtypeRank} is the mapping $\{
      \mathrm{\tt int8}\mapsto 1; \mathrm{\tt int32}\mapsto 2;
      \mathrm{\tt int64}\mapsto 3;
      \mathrm{\tt float16}\mapsto 4; \mathrm{\tt float32}\mapsto 5;
      \mathrm{\tt float64}\mapsto 6;
      \mathrm{\tt complex64}\mapsto 7; \mathrm{\tt complex128}\mapsto 8
    \}$.

  \item We introduce edges to record the difference between the dimension
    nodes. We denote this as $E_{\mathrm{dim ranks}}$.
    \begin{equation}
      \label{eq:to_induced_graph_dim_ranks}
      E_{\mathrm{dim ranks}} = \left\{(\iota^{-1}_{\mathrm{dim}}\left(d_i\right),
        \iota^{-1}_{\mathrm{dim}}\left(d_j\right)): d_j > d_i,
      d_i\in\operatorname{Dims}(e), d_j\in\operatorname{Dims}(e)\right\}
    \end{equation}

\end{itemize}

Based on these edges, we initialize the adjacency matrix as:
\begin{equation}\label{eq:directed_induced_graph_adj_matrix}
  A_{i j} =
  \begin{cases}
    1 & \mathrm{if}\; (i,j) \in E_{\mathrm{access,in-to-arg}}\\
    1 & \mathrm{if}\; (i,j) \in E_{\mathrm{output-to-access,in}}\\
    1 & \mathrm{if}\; (i,j) \in E_{\mathrm{arg-pos-to-access,in}}\\
    1 & \mathrm{if}\; (i,j) \in E_{\mathrm{index-to-access,in}}\\
    1 & \mathrm{if}\; (i,j) \in E_{\mathrm{dim-to-access,in}}\\
    1 & \mathrm{if}\; (i,j) \in E_{\mathrm{index-to-access,out}}\\
    1 & \mathrm{if}\; (i,j) \in E_{\mathrm{dim-to-access,out}}\\
    1 & \mathrm{if}\; (i,j) \in E_{\mathrm{length}}\\
    1 & \mathrm{if}\; (i,j) \in E_{\mathrm{dtype}}\\
    1 & \mathrm{if}\; (i,j) \in E_{\mathrm{length ranks}}\\
    1 & \mathrm{if}\; (i,j) \in E_{\mathrm{dtype ranks}}\\
    1 & \mathrm{if}\; (i,j) \in E_{\mathrm{dim ranks}}\\
    0 & \mathrm{otherwise}
  \end{cases}
\end{equation}

We define a node coloring based on the type of the node batched einsum entity
associated with it.

\begin{equation}\label{eq:directed_induced_graph_coloring}
  c_{i} =
  \begin{cases}
    1 & \mathrm{if}\; i \leq N_{\mathrm{arg}}\\

    2 & \mathrm{if}\; 0 < \left(i-N_{\mathrm{arg}}\right) \leq N_{\mathrm{index}}\\

    3 & \mathrm{if}\; 0 <
    \left(i-N_{\mathrm{arg}}-N_{\mathrm{index}}\right)\leq N_{\mathrm{access,in}}\\

    4 & \mathrm{if}\; 0 <
    \left(i-N_{\mathrm{arg}}-N_{\mathrm{index}}-N_{\mathrm{access,in}}\right)\leq
    N_{\mathrm{access,out}}\\

    5 & \mathrm{if}\; 0 <
    \left(i-N_{\mathrm{arg}}-N_{\mathrm{index}}-N_{\mathrm{access,in}}-N_{\mathrm{access,out}}\right)\leq
    b\\

    6 &\mathrm{if}\;  0 <
    \left(i-N_{\mathrm{arg}}-N_{\mathrm{index}}-N_{\mathrm{access,in}}-N_{\mathrm{access,out}}-b\right)\leq
    n\\

    7 &\mathrm{if}\;  0 <
    \left(i-N_{\mathrm{arg}}-N_{\mathrm{index}}-N_{\mathrm{access,in}}-N_{\mathrm{access,out}}-b-n\right)\leq
    N_{\mathrm{dtype}} \\

    8 &\mathrm{if}\;  0 <
    \left(i-N_{\mathrm{arg}}-N_{\mathrm{index}}-N_{\mathrm{access,in}}-N_{\mathrm{access,out}}-b-n-N_{\mathrm{dtype}}\right)\leq
    N_{\mathrm{length}}\\

    9 &\mathrm{if}\;  0 <
    \left(i-N_{\mathrm{arg}}-N_{\mathrm{index}}-N_{\mathrm{access,in}}-N_{\mathrm{access,out}}-b-n-N_{\mathrm{dtype}}-N_{\mathrm{length}}\right)\leq
    N_{\mathrm{dim}}
  \end{cases}
\end{equation}

Using the construction outlined above, we obtain the induced graph
associated with the batched einsum $e$ as $G$. This graph is defined by the
tuple $\left(N^{IDG}, \iota_{\mathrm{dtype}}, \iota_{\mathrm{lengths}}, A,
c\right)$, which are defined in
Equations~\eqref{eq:directed_induced_graph_n_idg},~\eqref{eq:directed_induced_graph_iota_dtype},~\eqref{eq:directed_induced_graph_iota_length},~\eqref{eq:directed_induced_graph_adj_matrix},
and~\eqref{eq:directed_induced_graph_coloring}, respectively.

\begin{example}[Induced Graph]\label{ex:directed_induced_graph}
  For the einsum $\Bigl(1$, $2$, $\left(i\right)$, $\left(\left(i k\right),
  \left(i j\right)\right)$, $\left(\left(A, B\right)\right)\Bigr)$, with
  $\operatorname{Shape}(A)=\operatorname{Shape}(B) = (72, 18)$ and
  $\operatorname{Dtype}(A)=\operatorname{Dtype}(B) = \mathrm{F64}$, we
  obtain the
  induced graph as shown in Figure~\ref{fig:demo_induced_graph}. It is worth
  noting that the nodes corresponding to $j$ and $k$ form a part of an
  automorphism in the graph.

  \begin{figure}
    \centering
    \includegraphics[width=0.6\textwidth]{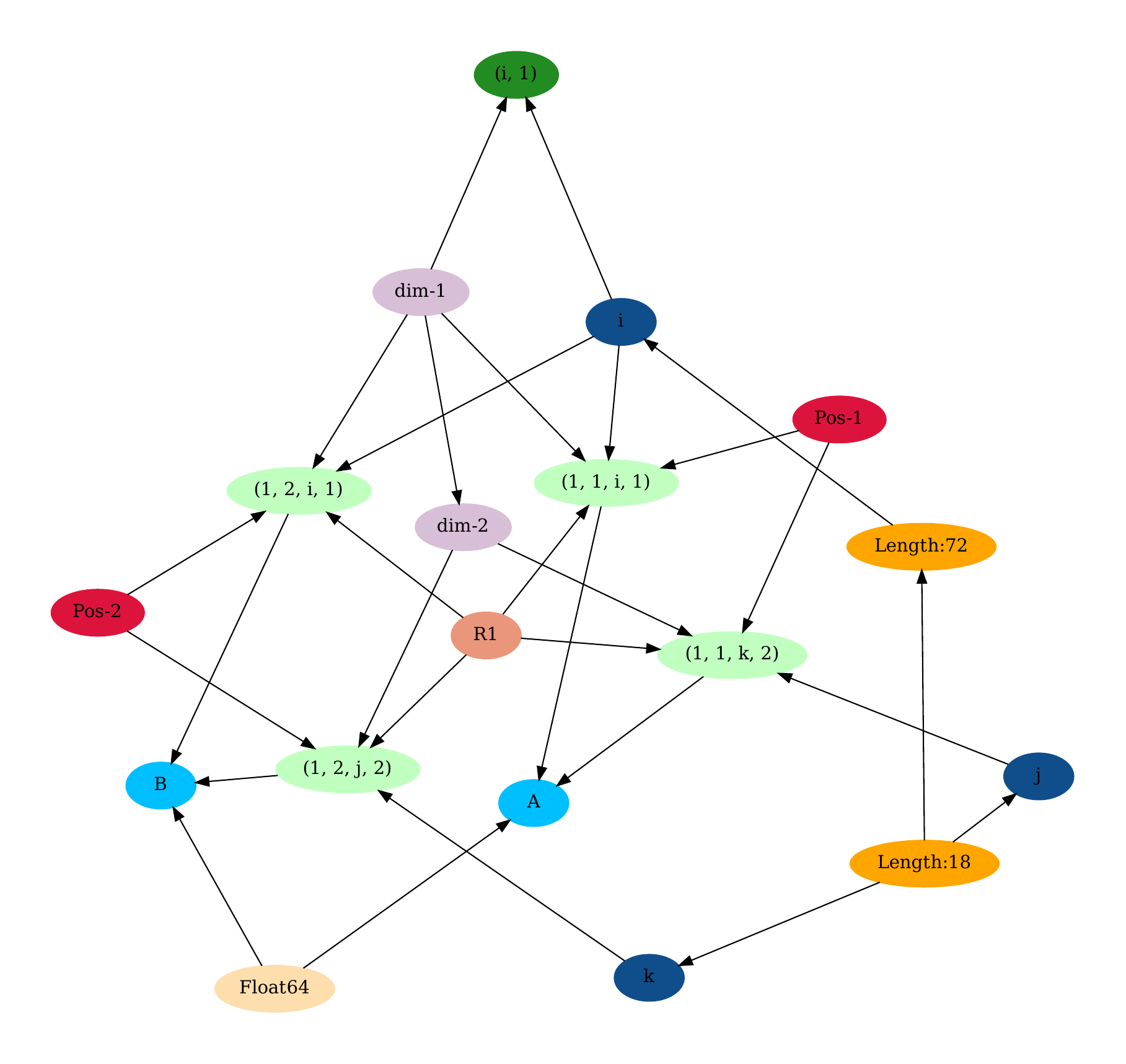}
    \caption{Induced graph for the batched einsum of
    Example~\ref{ex:directed_induced_graph}.}\label{fig:demo_induced_graph}
  \end{figure}
\end{example}

\subsection{Reconstructing a Batched Einsum from an Induced
Graph}\label{sec:from_directed_induced_graph}
In this section, we provide a mapping from $G$, an \textit{induced
graph} (Definition~\ref{defn:induced_directed_graph}), to a batched einsum.
We will consider that $G$ is defined by the tuple
$\bigl(N^{\mathrm{IDG}}$, $\iota_{\mathrm{dtype}}$, $\iota_{\mathrm{lengths}}$,
$A^{N^{\mathrm{IDG}} \times N^{\mathrm{IDG}}}$, $c_{N^{\mathrm{IDG}}}\bigr)$ and
provide a construction of a batched einsum. The aim of this section is to
invert the construction of Section~\ref{sec:to_directed_induced_graph}.

Before discussing the construction process, it is important to note that not
all induced graphs can be mapped to a valid batched einsum. An example
of a non-compliant graph is one that contains a node with a color value of 12,
as it violates the condition specified in
Equation~\eqref{eq:directed_induced_graph_coloring}.

We now introduce some terms that will be instrumental in our proposed
construction.

\begin{itemize}[label=$-$]
  \item We invert the coloring mapping of
    Equation~\eqref{eq:directed_induced_graph_coloring}, to obtain the
    following sets:
    \begin{equation}
      \begin{array}{ccc}
        V_{\mathrm{arg}} & = & \left\{i: i\in\{1, \ldots,
        N^{\mathrm{IDG}}\}, c_i = 1\right\}\\
        V_{\mathrm{index}} & = & \left\{i: i\in\{1, \ldots,
        N^{\mathrm{IDG}}\}, c_i = 2\right\}\\
        V_{\mathrm{access,in}} & = & \left\{i: i\in\{1, \ldots,
        N^{\mathrm{IDG}}\}, c_i = 3\right\}\\
        V_{\mathrm{access,out}} & = & \left\{i: i\in\{1, \ldots,
        N^{\mathrm{IDG}}\}, c_i = 4\right\}\\
        V_{\mathrm{output}} & = & \left\{i: i\in\{1, \ldots,
        N^{\mathrm{IDG}}\}, c_i = 5\right\}\\
        V_{\mathrm{arg-pos}} & = & \left\{i: i\in\{1, \ldots,
        N^{\mathrm{IDG}}\}, c_i = 6\right\}\\
        V_{\mathrm{dtype}} & = & \left\{i: i\in\{1, \ldots,
        N^{\mathrm{IDG}}\}, c_i = 7\right\}\\
        V_{\mathrm{length}} & = & \left\{i: i\in\{1, \ldots,
        N^{\mathrm{IDG}}\}, c_i = 8\right\}\\
        V_{\mathrm{dims}} & = & \left\{i: i\in\{1, \ldots,
        N^{\mathrm{IDG}}\}, c_i = 9\right\}\\
      \end{array}
    \end{equation}

  \item We define the neighbor relations in the graph $G$, as:
    \begin{equation}
      \begin{array}{ccc}
        \operatorname{Succs}(i) & = & \left\{j: j\in\{1, \ldots,
        N^{\mathrm{IDG}}\}, A_{i j} = 1\right\}\\
        \operatorname{Preds}(i) & = & \left\{j: j\in\{1, \ldots,
        N^{\mathrm{IDG}}\}, A_{j i} = 1\right\}\\
      \end{array}
    \end{equation}

\end{itemize}

The conditions required to invert the construction of
Section~\ref{sec:to_directed_induced_graph} consist of:

\begin{enumerate}[label={(C\arabic*)}, leftmargin=5em]
  \item For all $i\in\{1,\ldots,n\}$, $c_i$ must belong to $\{1,\ldots,
    9\}$.\label{prop:list_start}

  \item An argument node must not have any successors
    \[
      i \in V_{\mathrm{arg}} \Rightarrow \operatorname{Succs}(i) =
      \emptyset.
    \]

  \item A predecessor of an argument node must either be an input access node
    or a data type node
    \[
      i \in V_{\mathrm{arg}} \Rightarrow \operatorname{Preds}(i)
      \subseteq (V_{\mathrm{access,in}} \cup V_{\mathrm{dtype}}).
    \]

  \item An argument node must have exactly one data type node as predecessor
    \[
      n\in V_{\mathrm{arg}} \Rightarrow
      |\operatorname{Preds}(n) \cap V_{\mathrm{dtype}}| = 1.
    \]
    \label{prop:single_dtype_per_arg}

  \item A data type node can have only other data type nodes as predecessors
    \[
      i \in V_{\mathrm{dtype}} \Rightarrow \operatorname{Preds}(i) \subset
      V_{\mathrm{dtype}}.
    \]

  \item A successor of a data type node must either be an argument node
    or a data type node
    \[
      i \in V_{\mathrm{dtype}} \Rightarrow \operatorname{Succs}(i)
      \subseteq V_{\mathrm{arg}} \cup V_{\mathrm{dtype}}.
    \]

  \item A successor of an input access node must be an argument node
    \[
      i \in V_{\mathrm{access,in}} \Rightarrow \operatorname{Succs}(i)
      \subseteq V_{\mathrm{arg}}.
    \]

  \item An input access node must have exactly four predecessors, an index node, a
    dimension node, an argument position node and an output, i.e.
    \begin{equation}
      \label{eq:input_acc_node_four_preds}
      \begin{array}{lcc}
        i \in V_{\mathrm{access,in}} & \Rightarrow &
        |\operatorname{Preds}(i)| = 4\\
        i \in V_{\mathrm{access,in}} & \Rightarrow &
        |\operatorname{Preds}(i)\cap V_{\mathrm{arg-pos}}| = 1\\
        i \in V_{\mathrm{access,in}} & \Rightarrow &
        |\operatorname{Preds}(i)\cap V_{\mathrm{index}}| = 1\\
        i \in V_{\mathrm{access,in}} & \Rightarrow &
        |\operatorname{Preds}(i)\cap V_{\mathrm{output}}| = 1\\
        i \in V_{\mathrm{access,in}} & \Rightarrow &
        |\operatorname{Preds}(i)\cap V_{\mathrm{dim}}| = 1\\
      \end{array}
    \end{equation}

  \item A successor of an output node must be an input access node
    \[
      i \in V_{\mathrm{output}} \Rightarrow \operatorname{Succs}(i)
      \subseteq V_{\mathrm{access,in}}.
    \]

  \item An output node must not have any predecessors
    \[
      i \in V_{\mathrm{output}} \Rightarrow \operatorname{Preds}(i) =
      \emptyset.
    \]

  \item An output access node must not have any successors.
    \[
      i\in V_{\mathrm{access,out}} \Rightarrow \operatorname{Succs}(i) =
      \emptyset
    \]

  \item An output access node must have exactly two predecessors, an index
    node, and, a
    dimension node
    \begin{equation}
      \label{eq:output_acc_node_two_preds}
      \begin{array}{lcc}
        i \in V_{\mathrm{access,out}} & \Rightarrow &
        |\operatorname{Preds}(i)| = 2\\
        i \in V_{\mathrm{access,out}} & \Rightarrow &
        |\operatorname{Preds}(i)\cap V_{\mathrm{index}}| = 1\\
        i \in V_{\mathrm{access,in}} & \Rightarrow &
        |\operatorname{Preds}(i)\cap V_{\mathrm{dim}}| = 1\\
      \end{array}
    \end{equation}

  \item A successor of an index node must either be an input access node or
    an output access node
    \[
      i \in V_{\mathrm{index}} \Rightarrow \operatorname{Succs}(i)
      \subseteq \left(V_{\mathrm{access,in}}\cup V_{\mathrm{access,out}}\right).
    \]
  \item An index node must have exactly one node as predecessor. Furthermore, that
    predecessor must be an index length node.
    \[
      i \in V_{\mathrm{index}} \Rightarrow |\operatorname{Preds}(i)| = 1
      \; \text{and}\; |\operatorname{Preds}(i)
      \cap V_{\mathrm{length}}| = 1.
    \]
    \label{prop:single_length_per_index}

  \item A predecessor of an axis length node must be an axis length node
    \[
      i \in V_{\mathrm{length}} \Rightarrow \operatorname{Preds}(i)
      \subseteq V_{\mathrm{length}}
    \]

  \item A successor of an axis length node must either be an index node or
    an axis length node
    \[
      i \in V_{\mathrm{length}} \Rightarrow \operatorname{Succs}(i)
      \subseteq \left(V_{\mathrm{index}} \cup V_{\mathrm{length}} \right).
    \]

  \item A predecessor of a dimension node must be a dimension node
    \[
      i \in V_{\mathrm{dim}} \Rightarrow \operatorname{Preds}(i)
      \subseteq V_{\mathrm{dim}}.
    \]

  \item A successor of a dimension node must either be an access node or a
    dimension node
    \[
      i \in V_{\mathrm{dim}} \Rightarrow \operatorname{Succs}(i)
      \subseteq \left(V_{\mathrm{access,in}}\cup V_{\mathrm{access,out}} \cup
      V_{\mathrm{dim}}\right).
    \]

  \item An argument position node must not have any predecessors
    \[
      i \in V_{\mathrm{arg-pos}} \Rightarrow \operatorname{Preds}(i) =
      \emptyset.
    \]

  \item A successor of an argument position node must be an input access node
    \[
      i \in V_{\mathrm{arg-pos}} \Rightarrow \operatorname{Succs}(i)
      \subseteq V_{\mathrm{access,in}}.
    \]

  \item The subgraph of $G$ obtained by retaining only the dimension nodes and the
    corresponding edges must be a transitive tournament.
    Equivalently, there exists an
    ordering $v_1,\dots,v_{n}$ of the vertices in $V_{\text{dim}}$ such that
    \[
      v_j \in \operatorname{Succs}(v_i) \; \text{if and only if} \; i<j.
    \]

    These conditions mirror the edges outlined in the previous
    section, which correspond to the ranking of nodes across different
    dimensions as specified in~\eqref{eq:to_induced_graph_dim_ranks}.

  \item The subgraph of $G$ obtained by retaining only the length nodes and the
    corresponding edges must be a transitive tournament. Equivalently, there exists an
    ordering $v_1,\dots,v_{n}$ of the vertices in $V_{\text{length}}$ such that
    \[
      v_j \in \operatorname{Succs}(v_i) \; \text{if and only if} \; i<j.
    \]

    These conditions capture the edges outlined in the previous section, which
    correspond to the ranking of nodes across axis lengths as specified
    in~\eqref{eq:to_induced_graph_length_ranks}.

  \item The subgraph of $G$ obtained by retaining only the data type nodes and the
    corresponding edges must be a transitive tournament. Equivalently, there exists an
    ordering $v_1,\dots,v_{n}$ of the vertices in $V_{\text{dtype}}$ such that
    \[
      v_j \in \operatorname{Succs}(v_i) \; \text{if and only if} \; i<j.
    \]

    These conditions capture the edges outlined in the previous
    section, which correspond to the ranking of nodes across different
    numeric data types as specified in~\eqref{eq:to_induced_graph_dtype_ranks}.

  \item If
    $n_1 \in V_{\mathrm{access, in}}$, $n_2 \in V_{\mathrm{access, in}}$, $i_1 \in
    \operatorname{Preds}(n_1) \cap V_{\mathrm{index}}$,\\ $i_2 \in
    \operatorname{Preds}(n_2) \cap V_{\mathrm{index}}$,
    $\bigl(\operatorname{Succs}(n_1) \cap V_{\mathrm{arg}}$ $=$
    $\operatorname{Succs}(n_2) \cap V_{\mathrm{arg}}\bigr)$ $\wedge$
    $\left(\operatorname{Preds}(n_1) \cap V_{\mathrm{dim}} =
    \operatorname{Preds}(n_2) \cap V_{\mathrm{dim}}\right)$, then\\
    $\operatorname{Preds}(i_1)$ must be equal to
    $\operatorname{Preds}(i_2)$.
    This enforces that the arrays being of consistent shape by the following
    condition.

  \item If $n \in V_{\mathrm{access,out}}$, $i\in \operatorname{Preds}(n)
    \cap V_{\mathrm{index}}$, then $\operatorname{Succs}(i) \cap
    V_{\mathrm{access, in}}$ must be non-empty. This enforces that if an index
    node is indexing an output node then the index node is seen in one of the
    input nodes as well.

  \item If $n\in V_{\mathrm{access,in}}$, $i\in \operatorname{Preds}(n)
    \cap V_{\mathrm{index}}$, $d\in \operatorname{Preds}(n)
    \cap V_{\mathrm{dim}}$, $k \in \operatorname{Preds}(n)$ and $k
    \in V_{\mathrm{arg-pos}}$, then for all $o \in V_{\mathrm{output}}$,
    the set $\left\{n'\in V_{\mathrm{access,in}}: \left\{i, d, k,
      o\right\} \subseteq
    \operatorname{Preds}(n')\right\}$
    must be singleton. This condition enforces that the input indexing
    must be consistent for all outputs.

  \item The set $\left\{\operatorname{Preds}(n)\cap V_{\mathrm{index}}:
    n \in V_{\mathrm{access,out}}\right\}$ must have dimensionality of
    $|V_{\mathrm{access,out}}|$. This condition enforces that all output indices
    are distinct.

  \item For all $k\in V_{\mathrm{arg-pos}}$, $o \in V_{\mathrm{output}}$,
    $\mathrm{Acc}_{k,o} = \bigl\{n: k\in\operatorname{Preds}(n)$,
      $o\in\operatorname{Preds}(n)$, $n\in V_{\mathrm{access, in}}
    \bigr\}$. Then, the set $\left(\bigcup\limits_{a\in \mathrm{Acc}_{k,o}}
    \operatorname{Succs}(a)\right)$ must be singleton.

    This condition stipulates that only a single argument is indexed for an
    output.\label{prop:single_arg_per_row_col}

  \item For all $k\in V_{\mathrm{arg-pos}}$, $o \in V_{\mathrm{output}}$,
    if $\mathrm{Acc}_{k,o} = \{n: n\in V_{\mathrm{access, in}}$,
    $k\in\operatorname{Preds}(n)$, $o\in\operatorname{Preds}(n)\}$,
    and $\mathrm{Dims}_{k,o} =
    \bigcup\limits_{n\in\mathrm{Acc}_{k,o}}\operatorname{Preds}(n)\cap
    V_{\mathrm{dim}}$, then, the following conditions must hold:
    \begin{eqnarray*}
      |\{d: d \in \mathrm{Dims}_{k,o}, \operatorname{Preds}(d) \cap
      V_{\mathrm{dim}} = \emptyset\}| = 1\\
      |\{d: d \in \mathrm{Dims}_{k,o}, \operatorname{Preds}(d) \cap
      \mathrm{Dims}_{k,o} \neq \emptyset\}| = |\mathrm{Dims}_{k,o}| - 1
    \end{eqnarray*}
    This condition enforces that all consecutive dimensions of an argument are
    indexed.

  \item We define $\mathrm{Dims}_{\mathrm{out}} =
    \bigcup\limits_{n\in V_{\mathrm{{access, out}}}}\operatorname{Preds}(n)\cap
    V_{\mathrm{dim}}$. Then,
    \begin{eqnarray*}
      |\{d: d \in \mathrm{Dims}_{\mathrm{out}}, \operatorname{Preds}(d) \cap
      V_{\mathrm{dim}} = \emptyset\}| = 1\\
      |\{d: d \in \mathrm{Dims}_{\mathrm{out}}, \operatorname{Preds}(d) \cap
      \mathrm{Dims}_{\mathrm{out}} \neq \emptyset\}| =
      |\mathrm{Dims}_{\mathrm{out}}| - 1
    \end{eqnarray*}
    This condition enforces that all consecutive dimensions of the output are
    indexed.

    \label{prop:list_end}
\end{enumerate}

\begin{remark}
  Our induced graph construction from
  Section~\ref{sec:to_directed_induced_graph} satisfies the
  properties~\ref{prop:list_start}--\ref{prop:list_end}.
\end{remark}

For an induced graph that satisfies the compliance properties mentioned
above, we can obtain its corresponding batched einsum by following these steps:
\begin{enumerate}[label={\textbf{Step \arabic*}.},
    leftmargin=5em]
  \item Choose two functions $\operatorname{IndexName}: \mathbb{Z}^{+}
    \to \mathrm{identifier}$, for naming an index in the reconstructed batched
    einsum and $\operatorname{ArgName}: \mathbb{Z}^{+} \to \text{identifier}$,
    for naming an argument in the reconstructed batched einsum.
  \item Define a mapping for argument ordering in the reconstructed
    batched einsum, denoted as $\iota_{\mathrm{arg-pos}}^{\mathrm{inferred}}:
    V_{\mathrm{arg-pos}} \mapsto \{1, \ldots, |V_{\mathrm{arg-pos}}|\}$ such that
    if $n_1, n_2 \in V_{\mathrm{dim}}$ and $n_1 < n_2$, then
    $\iota_{\mathrm{arg-pos}}^{\mathrm{inferred}}(n_1) <
    \iota_{\mathrm{arg-pos}}^{\mathrm{inferred}}(n_2)$.

  \item Define a mapping for the output ordering in the reconstructed batched
    einsum, denoted as $\iota_{\mathrm{output}}^{\mathrm{inferred}}:
    V_{\mathrm{output}} \mapsto \{1, \ldots, |V_{\mathrm{output}}|\}$ such that
    if $n_1, n_2 \in V_{\mathrm{output}}$ and $n_1 < n_2$, then
    $\iota_{\mathrm{output}}^{\mathrm{inferred}}(n_1) <
    \iota_{\mathrm{output}}^{\mathrm{inferred}}(n_2)$.

  \item Define a mapping for the index symbol ordering in the reconstructed
    batched einsum, denoted as $\iota_{\mathrm{index}}^{\mathrm{inferred}}:
    V_{\mathrm{index}} \mapsto \{1, \ldots, |V_{\mathrm{index}}|\}$ such that
    if $n_1, n_2 \in V_{\mathrm{index}}$ and $n_1 < n_2$, then
    $\iota_{\mathrm{index}}^{\mathrm{inferred}}(n_1) <
    \iota_{\mathrm{index}}^{\mathrm{inferred}}(n_2)$.\label{step:reconstruct_digraph_iota_index}

  \item Define a mapping for the argument ordering the reconstructed batched
    einsum, denoted as $\iota_{\mathrm{arg}}^{\mathrm{inferred}}:
    V_{\mathrm{arg}} \mapsto \{1, \ldots, |V_{\mathrm{arg}}|\}$ such that
    if $n_1, n_2 \in V_{\mathrm{arg}}$ and $n_1 < n_2$, then
    $\iota_{\mathrm{arg}}^{\mathrm{inferred}}(n_1) <
    \iota_{\mathrm{arg}}^{\mathrm{inferred}}(n_2)$.\label{step:reconstruct_digraph_iota_arg}

  \item Define the mapping for the dimensions in the reconstructed batched
    einsum, $\iota^{\mathrm{inferred}}_{\mathrm{dim}}:
    V_{\mathrm{dim}}\mapsto \{1, \ldots, |V_{\mathrm{dim}}|\}$, as:
    \begin{equation}
      \iota^{\mathrm{inferred}}_{\mathrm{dim}}(n) = 1 + |\operatorname{Preds}(n) \cap V_{\mathrm{dim}}|.
    \end{equation}

  \item Find the output index, denoted by $\mathcal{I}^{\mathrm{out,
    inferred}}$, as:
    \begin{equation}
      \mathcal{I}^{\mathrm{out, inferred}} =
      \operatorname{IndexName}\left(\iota_{\mathrm{index}}^{\mathrm{inferred}}\left(\operatorname{OutIndex}(k)\right)\right)
    \end{equation}
    where,
    \begin{equation}
      \label{eq:infer_out_index}
      \operatorname{OutIndex}(k) \in \left\{i:
        \left(\iota^{\mathrm{inferred}}_{\mathrm{dim}}\right)^{-1}(k) \in
        \operatorname{Preds}(n), n \in
      V_{\mathrm{access,out}}\right\}.
    \end{equation}

    From the criterion in equation~\eqref{eq:output_acc_node_two_preds}, we can
    see the there is exactly one solution for $\operatorname{OutIndex}(k)$ in
    the above equation~\eqref{eq:infer_out_index}.

  \item Find the argument dimensions, denoted by
    $\operatorname{ArgDim}(k)$, as:
    \begin{equation}
      \begin{array}{ll}
        \operatorname{ArgDim}(k) = \Bigl|\bigl\{  d\in V_{\mathrm{dim}}:& \{d,
          \left(\iota^{\mathrm{inferred}}_{\mathrm{arg-pos}}\right)^{-1}(k)\}\subset\operatorname{Preds}(a),\\
        &  a\in V_{\mathrm{access, in}}\bigr\}\Bigr|\\
      \end{array}
    \end{equation}

  \item Find the input index list, denoted by $\mathcal{I}^{\mathrm{in,
    inferred}}$, as:
    \begin{equation}
      \mathcal{I}^{\mathrm{in, inferred}} =
      \left(\left(\operatorname{IndexName}\left(\iota_{\mathrm{index}}^{\mathrm{inferred}}\left(\operatorname{InputIndex}(d,
      j)\right)\right)\right)_{d=1}^{\operatorname{ArgDim}(j)}\right)_{j=1}^{|V_{\mathrm{arg-pos}}|}
    \end{equation}
    where,
    \begin{equation}
      \label{eq:infer_in_index}
      \begin{split}
        \operatorname{InputIndex}(d, j) \in \bigl\{i: &i\in
          \operatorname{Preds}(n),\\
          & i\in V_{\mathrm{index}},\\
          &\{\left(\iota^{\mathrm{inferred}}_{\mathrm{arg-pos}}\right)^{-1}[j],\left(\iota^{\mathrm{inferred}}_{\mathrm{dim}}\right)^{-1}[d]\}
          \subset \operatorname{Preds}(n),\\
        & n \in V_{\mathrm{access,in}}\bigr\}
      \end{split}
    \end{equation}

    From the criterion in equation~\eqref{eq:input_acc_node_four_preds}, we can
    see the there is exactly one solution for $\operatorname{InputIndex}(d, j)$ in
    the above equation~\eqref{eq:infer_in_index}.

  \item Find the argument sequences as
    $\left(\left(\mathcal{A}^{i, j,
    \mathrm{inferred}}\right)_{j=1}^{|V_{\mathrm{arg-pos}}|}\right)_{i=1}^{|V_{\mathrm{output}}|}$,
    where $\mathcal{A}^{i, j, \mathrm{inferred}}$
    is an array with the symbol
    $\operatorname{ArgName}\left(\iota^{\mathrm{inferred}}_{\mathrm{array}}(\mathrm{ArgLabel}(i,
    j))\right)$, with the shape
    $\left(\iota_{\mathrm{length}}(\mathrm{LengthLabel}(i,
    j))\right)_{j=1}^{\mathrm{\textsc{ArgDim(j)}}}$ and with the data type
    $\iota_{\mathrm{dtype}}\left(\mathrm{DtypeLabel}(i, j)\right)$, where,

    \begin{equation}
      \label{eq:inferred_arg_label}
      \begin{split}
        \mathrm{ArgLabel}(i, j) \in \bigl\{a:
          & \{\left(\iota^{\mathrm{inferred}}_{\mathrm{output}}\right)^{-1}(i),\\
          & \left(\iota^{\mathrm{inferred}}_{\mathrm{arg-pos}}\right)^{-1}(j)\}
          \subset \operatorname{Preds}(n),\\
          & n\in\left(\operatorname{Preds}(n)\cap V_{\mathrm{access,
          in}}\right),\\
          & a \in V_{\mathrm{arg}}
        \bigr\}
      \end{split}
    \end{equation}
    \begin{equation}
      \label{eq:infer_dtype}
      \mathrm{DtypeLabel}(i, j) \in \left\{d: V_{\mathrm{dtype}},
        d\in\operatorname{Preds}(\mathrm{ArgLabel}(i,
      j))\right\}
    \end{equation}
    \begin{equation}
      \label{eq:infer_length}
      \mathrm{LengthLabel}(i, j) \in \left\{l: l\in V_{\mathrm{length}},
        l\in \operatorname{Preds}(\mathrm{InputIndex}(i, j))
      \right\}
    \end{equation}

    From the
    criteria~\ref{prop:single_arg_per_row_col},~\ref{prop:single_dtype_per_arg},and,~\ref{prop:single_length_per_index},
    we can see that the
    equations~\eqref{eq:inferred_arg_label},~\eqref{eq:infer_dtype},
    and,~\eqref{eq:infer_length} have exactly one solution.

  \item Using the previous steps, we reconstruct our batched einsum
    as a batched einsum defined by the tuple $\Bigl(|V_{\mathrm{output}}|$,
      $|V_{\mathrm{arg-pos}}|$, $\mathcal{I}^{\mathrm{out, inferred}}$,
      $\left(\mathcal{I}^{k, \mathrm{in,
      inferred}}\right)_{i=1}^{|V_{\mathrm{arg-pos}}|}$,
      $\left(\left(\mathcal{A}^{i, j,
    \mathrm{inferred}}\right)_{j=1}^{|V_{\mathrm{arg-pos}}|}\right)_{i=1}^{|V_{\mathrm{output}}|}\Bigr)$.
\end{enumerate}

\section{DG-FEM Batched Einsum Kernel Patterns}

\lstset{style=pythonstyle, frame=single, numbers=left,stepnumber=1,
morekeywords={INSTRUCTIONS,DOMAINS,ARGUMENTS, SUBSTITUTION, RULES,end}}
\begin{lstlisting}[caption={A kernel performing $b$-applications
                            of the face-mass operation. The
                            variables \texttt{F\_1}, $\ldots$,\texttt{F\_b}
                            correspond to the surface fluxes.
                            \texttt{jac} is the substitution rule corresponding
                            to the Jacobian terms and \texttt{mat} is the
                            rule corresponding to the tabulated reference
                            surface integral.},
                   label={lstng:feinsum_facemass_kernels}]
jac(f, e)    := J[f, e]
mat(i, f, j) := D[i, f, j]
f1(f, e, j)  := F_1[f, e, j]
f2(f, e, j)  := F_2[f, e, j]
# ...
fb(f, e, j)  := F_b[f, e, j]

for iel, idof
 y1[iel,idof] = sum([iface,jdof], f1(iface,iel,jdof)
                                  *jac(iface,jdof)
                                  *mat(idof,iface,jdof]))
 y2[iel,idof] = sum([iface,jdof], f2(iface,iel,jdof)
                                  *jac(iface,jdof)
                                  *mat(idof,iface,jdof]))
 # ...
 yb[iel,idof] = sum([iface,jdof], fb(iface,iel,jdof)
                                  *jac(iface,jdof)
                                  *mat(idof,iface,jdof]))
end iel, idof
\end{lstlisting}

\begin{lstlisting}[caption={A kernel performing $b$-applications
                            of the local divergence operation. The
                            tuples \{(\texttt{u1}, \texttt{v1}, \texttt{w1}),
                            $\ldots$, (\texttt{ub}, \texttt{vb}, \texttt{wb})\}
                            correspond to the individual components of the
                            input vector fields.
                            \texttt{jac} is the substitution rule corresponding
                            to the Jacobian terms and \texttt{mat} is the
                            reference derivative matrix.},
                   label={lstng:feinsum_divergence_kernels}]
jac(x,r,e) := J[x,r,e]
mat(r,i,j) := D[r,i,j]
f1(x,e,j)  := u1[e, j] if x == 0 else (v1[e, j] if x == 1 else w1[e, j])
f2(x,e,j)  := u2[e, j] if x == 0 else (v2[e, j] if x == 1 else w2[e, j])
# ...
fb(x,e,j)  := ub[e, j] if x == 0 else (vb[e, j] if x == 1 else wb[e, j])

for iel, idof
 y1[iel,idof] = sum([itopo_dim,iambient_dim,jdof], f1(itopo_dim,iel,jdof)
                                                   *jac(itopo_dim,iambient_dim,iel)
                                                   *mat(iambient_dim,idof,jdof]))
 y2[iel,idof] = sum([itopo_dim,iambient_dim,jdof], f2(itopo_dim,iel,jdof)
                                                   *jac(itopo_dim,iambient_dim,iel)
                                                   *mat(iambient_dim,idof,jdof]))
 # ...
 yb[iel,idof] = sum([itopo_dim,iambient_dim,jdof], fb(itopo_dim,iel,jdof)
                                                   *jac(itopo_dim,iambient_dim,iel)
                                                   *mat(iambient_dim,idof,jdof]))
end iel, idof
\end{lstlisting}

\begin{lstlisting}[caption={A kernel performing $b$-applications
                            of the local gradient operation. The
                            variables \{\texttt{u1}, $\ldots$, \texttt{ub}\},
                            correspond to the input scalar fields.
                            \texttt{jac} is the substitution rule corresponding
                            to the Jacobian terms and \texttt{mat} is the
                            reference derivative matrix.},
                   label={lstng:feinsum_gradient_kernels}]
jac(x,r,e) := J[x,r,e]
mat(r,i,j) := D[r,i,j]
f1(x,e,j)  := u1[e, j]
f2(x,e,j)  := u2[e, j]
# ...
fb(x,e,j)  := ub[e, j]

for iel, idof
 y1[itopo_dim, iel,idof] = sum([iambient_dim,jdof], f1(iel,jdof)
                                                   *jac(itopo_dim,iambient_dim,iel)
                                                   *mat(iambient_dim,idof,jdof]))
 y2[itopo_dim, iel,idof] = sum([iambient_dim,jdof], f2(iel,jdof)
                                                   *jac(itopo_dim,iambient_dim,iel)
                                                   *mat(iambient_dim,idof,jdof]))
 # ...
 yb[itopo_dim, iel,idof] = sum([iambient_dim,jdof], fb(iel,jdof)
                                                   *jac(itopo_dim,iambient_dim,iel)
                                                   *mat(iambient_dim,idof,jdof]))
end iel, idof
\end{lstlisting}

\end{document}